\documentclass[acmsmall]{acmart}
\usepackage{cancel}

\AtBeginDocument{%
  \providecommand\BibTeX{{%
    \normalfont B\kern-0.5em{\scshape i\kern-0.25em b}\kern-0.8em\TeX}}}

\setcopyright{acmcopyright}
\acmJournal{PACMHCI}
\acmArticle{202}

\setcopyright{acmlicensed}
\acmJournal{PACMHCI}
\acmYear{2020} \acmVolume{4} \acmNumber{ISS}\acmMonth{11} \acmPrice{15.00}\acmDOI{10.1145/3427330}

\usepackage{csvsimple}
\usepackage{float}
\usepackage[caption = false]{subfig}
\usepackage[]{graphicx}

\usepackage{xcolor,colortbl}
\usepackage[all]{nowidow}

\begin{document}

\clubpenalty=9996
\widowpenalty=9999
\brokenpenalty=4991
\predisplaypenalty=10000
\postdisplaypenalty=1549
\displaywidowpenalty=1602
\looseness=-5

\title[Understanding Gesture and Speech Multimodal Interactions]{Understanding Gesture and Speech Multimodal Interactions for Manipulation Tasks in Augmented Reality Using Unconstrained Elicitation}

\author{Adam S. Williams}
\email{AdamWil@colostate.edu}

\affiliation{%
  \institution{Colorado State University, USA}
  \streetaddress{2545 Research Blvd}
  \city{Fort Collins}
  \state{Colorado}
  \postcode{80526}
}

\author{Francisco R. Ortega}
\email{F.Ortega@colostate.edu}

\affiliation{%
  \institution{Colorado State University,  USA}
  \streetaddress{2545 Research Blvd}
  \city{Fort Collins}
  \state{Colorado}
  \postcode{80526}
}

\renewcommand{\shortauthors}{Adam S. Williams, et al.}

\begin{abstract}

This research establishes a better understanding of the syntax choices in speech interactions and of how speech, gesture, and multimodal gesture and speech interactions are produced by users in unconstrained object manipulation environments using augmented reality. The work presents a multimodal elicitation study conducted with 24 participants. The canonical referents for translation, rotation, and scale were used along with some abstract referents (create,  destroy,  and select). In this study time windows for gesture and speech multimodal interactions are developed using the start and stop times of gestures and speech as well as the stoke times for gestures. While gestures commonly precede speech by 810 \textcolor{black}{ms} we find that the stroke of the gesture is commonly within 100 ms of the start of speech. Indicating that the information content of a gesture and its co-occurring speech are well aligned to each other. Lastly, the trends across the most common proposals for each modality are examined. Showing that the disagreement between proposals is often caused by a variation of hand posture or syntax. Allowing us to present aliasing recommendations to increase the percentage of users' natural interactions captured by future multimodal interactive systems.


\end{abstract}



\newcommand{\AR}{$\mathcal{AR}\:$}
\newcommand{\CAR}{$\mathcal{CAR}\:$}
\newcommand{\VRD}{$V_{rd}\:$}
\newcommand{\MC}{$\mathcal{MC}\:$}
\newcommand{\CDR}{$\mathcal{CDR}\:$}

\maketitle

\section{Introduction}

Establishing impactful unimodal and multimodal interaction techniques for augmented reality (AR) head-mounted displays (HMDs) starts with understanding unconstrained user behavior. Gesture and speech show promise as the inputs that will be well suited for use in AR-HMDs. Both of these modalities can be tracked with the sensors that come standard on most consumer-available AR-HMDs such as the Microsoft Hololens 2. This minimalism is beneficial. When using AR-HMDs people will likely seek to carry as little extra technology as possible. 

Gestures and speech have strengths as both unimodal and multimodal inputs \cite{MOR12}. These strengths have not yet fully been examined. Speech has been found well suited for abstract tasks such as multi-object manipulation \cite{PIU+14} or selecting a device out of a set of devices \cite{VAT+15b}. Gestures have been found well suited for direct manipulation \cite{PIU+14}. The combination of these modalities can provide a more rich interaction environment than either alone. By understanding the strengths and synergies of these modalities we can better design systems for the end-user.

We can see some of the impacts of new interaction paradigms in the widespread use of multi-touch devices (e.g.,touch screen cell phones) reaching populations that do not commonly use computers but can benefit from the use of technology \cite{SUS+13}. Augmented reality is one of the technologies expected to become pervasive in the future, and with that, interactions in AR-HMD environments will become pervasive. Proof AR-HMDs' increased prevalence can been seen in the the United States government's purchase of $100,000$ Microsoft HoloLens 2 units for Army use \cite{DOD}. There is little standardization for mid-air gestures AR environments \cite{BIL07}, the same can be said for speech inputs. Co-occurring gesture and speech interactions, where both gestures and speech are used to convey a message within close temporal proximity of each other, have been analyzed within the context of human to human interaction \cite{MCN05, KEL+10, MOL+12}, however, the unconstrained generation of these inputs in human-computer interaction (HCI) has been far less commonly examined \cite{MIG+93, MOR12, IRA+06}.  

This research presents a study in which participants are tasked with interacting with a virtual object both unimodally and multimodally in an optical see-through AR-HMD environment. These interactions were unconstrained. Gestures, speech, and co-occurring gesture and speech interactions were each tested independently. The main goal of this research was to provide insight on speech interactions, with and without gestures, for object manipulation in AR. To provide robust comparisons, unimodal gesture alone interactions were also examined. 

The \textbf{contributions} of this research include a detailed analysis of these input modalities' interactions and insights into the changes in those interactions when used multimodally as opposed to unimodally are given. Instead of presenting a single consensus set for each modality, we highlight the common proposals, themes across proposals, and the syntax used for speech interactions. Lastly, timing windows based on the phases of a co-occurring gesture and speech interaction are constructed. Showing that the information content of an interaction is closely aligned with the stroke of a gesture. Based on those findings this paper establishes some guidelines for multimodal gesture and speech input development in this emerging area.

\subsection{Motivation}

Interactions with systems should be intuitive \cite{NIE+04}. One way of achieving that is by leveraging interaction modalities that we are familiar with. Interpersonal communication is rich with gesture and speech interaction \cite{MCN05}. Communication is formed in both gesture and speech channels simultaneously, with each channel impacting the formation of a message by the other channel \cite{KEL+10}. Enabling a system to accept gesture and speech as both unimodal and multimodal input channels, is an important step towards creating intuitive augmented reality interaction design. 

When participants were given the option to chose modalities, they chose to combine gesture and speech inputs  60\% to 70\% of the time \cite{COR+05, HAU89}. This preference can be used to improve recognition \cite{EIS+04}. End-users feel that interactions with a system are more natural when they can chose input modalities based on their preference \cite{KAR+18, BAI+18}. By leveraging this preference and multimodal inputs, many benefits can be realized. The use of multiple input channels can lead to mutual disambiguation of information lost in the other channel \cite{OVI00, KOO+93, KAI+03}, as well as lead to less verbose interactions by allowing for two communication channels to send non-redundant information simultaneously \cite{GOL+93}. Gesticulation is closely linked to the structure of co-occurring speech, allowing for better error recovery in recognizers \cite{KOO+93}.

Optical see-trough AR-HMDs (e.g., Magic Leap One and Microsoft Hololens versions 1 \& 2) are starting to implement gesture and speech interactions. That said, these interactions could still use much improvement. Some of the interactions implemented seem built to improve recognition accuracy rather than improving user experience. For example, Magic Leap's C gesture is fairly easy to detect (being a static symbolic gesture) but may not be the most intuitive. Often if gesture sets are not designed with an emphasis on recognition they are designed by experts \cite{WOL+11}. User-defined gesture sets have been shown to be up to 24\% more memorable \cite{NAC+13} and to be preferred to expert-designed gesture sets \cite{WOB+09}.

This work is not on multimodal fusion (or recognition) \cite{CAR+06}, rather, it is on multimodal interaction, input generation, and design. Nevertheless, the results of our study can be used by researchers working on multimodal fusion. We use participatory design guidelines to work with potential end-users of AR-HMDs to find what inputs within each modality they would instinctively use \cite{WOB+09, MOR12}. The timing information for phases of a multimodal interaction can help tune recognition windows in multimodal fusion systems. The combination of work on elicitation, such as this study, and multimodal fusion will help HCI build systems with more natural interactions. The technological gap between the feasibility of traditional inputs and gesture with speech inputs is being minimized, soon the later may become more efficient \cite{BAI+18}. This work provides information on the top few interaction proposals for each modality, interaction themes across modalities, co-occurring gesture and speech timing information by phase of interaction, and design guidelines on input design for AR building environments.

\section{Previous Work}

\subsection{Gesture Elicitation} 

Elicitation is a type of study that aims at mapping inputs to emerging technologies through participatory design. The elicited inputs should be discoverable to novice users of systems \cite{WOB+09}. A second product of elicitation studies is a better understanding of user behavior. Elicitation studies have shown that upper-body gestures are preferred in whole-body gesture systems \cite{ORT+19}, and that gestures produced are impacted by the size of the object \cite{TAR+18, PLA+17}. Elicitation has seen use for many input domains such as multi-touch surfaces \cite{MIC+09, BUC+13}, and mobile devices \cite{RUI+11}, to internet of things  use \cite{VAT+15b}. 

Elicitation studies typically use a Wizard of Oz (WoZ) experiment design \cite{WOB+05, WOB+09}. WoZ experiment design can be used to remove the gulf of execution between the participant and the system by removing the systems input recognizer \cite{WOB+09}. In a WoZ elicitation experiment, a participant is shown a command (referent) to execute such as \textit{move down}. The participant generates an input proposal for that referent which causes an experimenter to emulate the recognition of that input. In this work that is changed slightly to allow for better collection of speech results. For the command \textit{move down} in this experiment, a participant was shown a virtual object moving down after which they would be asked to generate a command to produce that effect. By running the study this way we were able to collect inputs for a system that does not have a perfect recognizer or fusion model.  


One outcome of an elicitation study is the production of a mapped set of inputs called a consensus set \cite{COH+12, BUC+13}. More useful than a single set of mapped inputs is the observational data that comes from elicitation studies. This includes insight on the formation of inputs, the times surrounding input generation, and trends in user preferences for inputs and input modalities. An example of these extended benefits is the finding that the size of a gesture proposed is impacted by the size of the object shown \cite{TAR+18}. This work extends previous gesture elicitation studies in AR \cite{PIU+13} by testing the additional modalities of speech alone and multimodal gesture and speech interactions and allowing unconstrained gesture proposals for each referent. Furthermore, the set of interactions presented here shows the top few proposals allowing better interpretation of trends in gesture formation. 

\subsection{Gesture and Speech studies}

A large portion of multimodal gesture and speech input studies have been focused on finding ways to combine them using multimodal fusion models \cite{BOL80, CAR+06, JOH+97, OVI+97}. There has also been work on finding the timing windows for co-occurring gesture and speech interactions \cite{LEE+08}. Some of this work looks at the usability of constrained sets of inputs such as limited gesture sets \cite{CHA+16} or limited speech dictionaries \cite{LEE+08}. These types of works look for a better understanding of a combination of the feasibility of inputs, the adaptability of people to constrained inputs, and the implementation or accuracy of fusion models for gesture and speech recognition. These works typically start with live mapped inputs and test usability or accuracy. \textbf{The work presented here is very different in that there are no constraints imposed on input proposals, and deliberate efforts were made to remove text based priming in the speech condition}. Participants are invited to generate any input proposal they see fit for the given referent and input modality.

While a few studies look at gesture and speech inputs have examined mid-air gestures \cite{HAU+93, CAR+06, LEE+08, MOR12, ANA+12, KHA+19}, some only looked at a subset of gesturing such as pointing gestures \cite{BOU+98, ROB98}, paddling gestures \cite{IRA+06}, or two dimensional (2D) gestures \cite{MIG+93, ROB98}. The work presented here examines any mid-air gesture and / or utterance that a participant feels is appropriate for a given referent.

This study extends previous works done on multimodal gesture and speech elicitation \cite{KHA+19, MOR12}. This extension is seen in the results reported and the methodology used. A previous study on interactions for computer-aided design program usage on 2d screens tested both gesture and gesture or speech interactions \cite{KHA+19}. In that experiment, gestures were tested independently then gesture with optional speech was tested. This is different from our choice to examine each input individually. In both studies the referents were shown as animations, however, in this study participants were told that they were interacting with a system whereas Khan et al. asked participants to describe the referents to another person via a video chat \cite{KHA+19}. The use case of computer-aided design as well as the choice of observing interactions compared to referent descriptions is markedly different, with examples of the referents used there being \textit{extrude surface} or \textit{pan}.

This work also extends the results of a study done on eliciting commands for television-based web browsing \cite{MOR12}. That study used paired elicitation where participants would sit in groups on a couch and propose either gesture, speech, or gesture and speech commands, as compared to the individual elicitation technique used here. That study also only examined the input modalities in a single pass where participants were allowed to produce any command in any modality or a combination of modalities. An important distinction is that referents were shown as text and read aloud by the experimenter in Morris, 2012 \textcolor{black}{\cite{MOR12}}. In this study we examine interaction proposals without text prompting.



This work differs from previous gesture and speech elicitation studies in several important ways. This work does not present users with any text when showing referents. Participants are not paired and are asked to produce an input for each modality. This is in comparison to prior works which commonly allows users to chose which modality they use when generating input proposals \cite{VIL+20}. This work aims on finding intuitive inputs across the gesture, speech, and co-occurring gesture and speech interactions. This work does not attempt to improve gesture or speech recognition, nor does it attempt to build better multimodal fusion models. It is our hope that these results can be used towards those goals in future studies.

\section{Methods}

\subsection{Pilot Studies}

Two versions of this study were run to assess the impact of referent display on proposal generation. The results of these pilot studies were used to inform the methodology decisions made in this experiment. These each used 6 people. In one of the pilot studies, we display the referents as text on the screen, which is different from our final design. The first pilot study's design is comparable to \cite{MOR12, WOB+09}. In the second pilot study, we displayed the referent by showing the participants an animation of the intended effect of the interaction they would propose. The second pilot study's design is comparable to \cite{KHA+19}. Both the pilot studies and this study tested the same input modalities, those being, gesture and speech, speech alone, and gesture alone.

In the first pilot study, there was evidence that text referents primed speech production. If the referent was \textit{move right} the utterance was commonly ``move right''. This effect was more pronounced for translations, rotations had more variance in proposals but still showed signs of biasing. Repeating referents when producing speech proposals, such as saying ``new tab'' for the referent \textit{new tab}, can be seen in the results of Morris, 2012 \cite{MOR12}. When the referents were shown as animations in the second pilot study, people would often mirror that animation in the gesture they produced. These mirrored gestures were often direct manipulations which are not uncommon in gesture interfaces \cite{3duibook}, however, when designing inputs that priming could be problematic. The effect animations biasing gestures can be seen in the study done by Khan et al. 2019 \cite{KHA+19}, such as a \textit{pan} gesture that mirrors the motion of the animation used.

This study's goal was to understand user speech behavior both alone and when co-occurring with gestures. With that in mind, we have chosen to show the referents as an animation. The only text shown to the participant was the input modality requested (e.g. ``gesture only'', ``speech only'', ``gesture and speech''). This will allow us to have more robust speech results than when showing a text based referent. Another choice in elicitation methodology used in this experiment was to not have think aloud protocol as seen in \cite{WOB+09}. The process of thinking out loud while generating speech proposals would confound the results, making speech data less reliable. 

\subsection{Methodology}

\begin{figure}[htbp]
    \centering
      \includegraphics[width=.70\columnwidth, keepaspectratio]{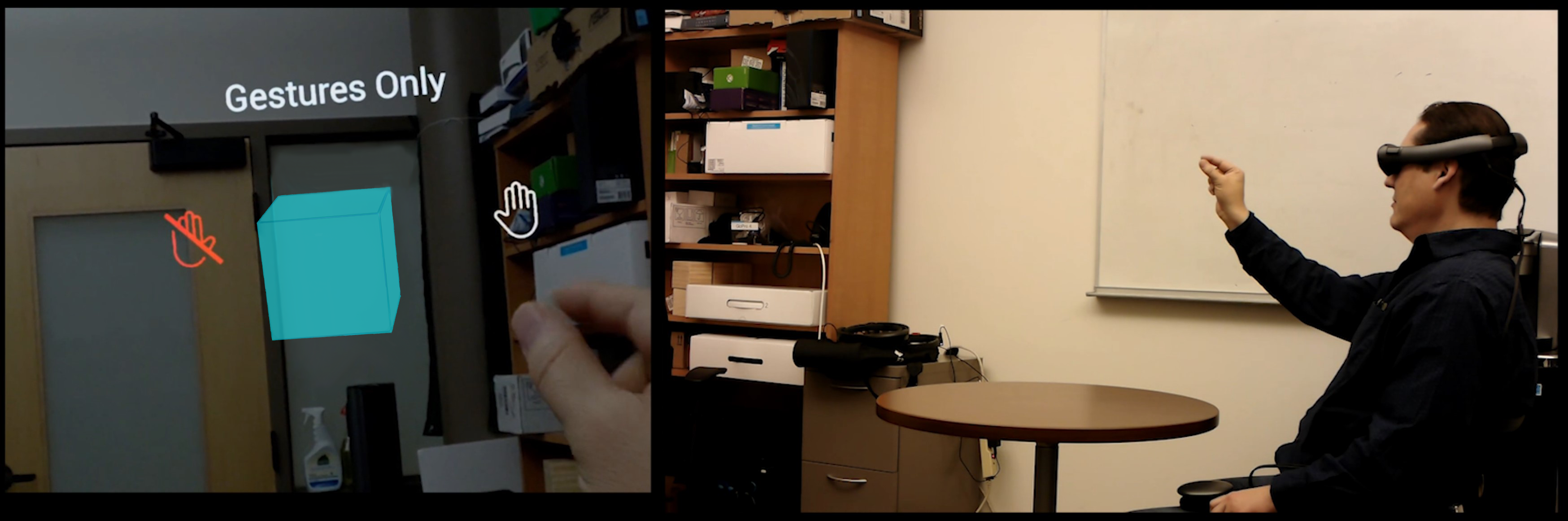}
    \caption{Experimental Set up: Left, participant view, Right: participant}
    \label{fig:trialExample}
\end{figure}

This study was run as a within-subjects (i.e. repeated measures) elicitation study. The goal of this work was to gain a better understanding of the production of gestures, speech, and co-occurring gestures and speech when interacting with three-dimensional (3D) objects in an optical see-through AR-HMD. Participants were asked to generate proposals for gesture alone, speech alone, and multimodal gesture and speech interactions. These input modalities were presented in a counterbalanced order. Within each input, participants were asked to generate an interaction proposal for each referent. Meaning that a participant may be assigned the speech input modality first, then be asked to generate a speech proposal for each referent before progressing to either the gesture or gesture and speech condition. Referents were displayed in random order with each occurring once per input modality. The experimental setup is illustrated in Figure \ref{fig:trialExample}. Participants were told that they were guessing the interaction that someone in a different room was using to execute the referent they were presented with. A single referent sequence was a blank screen, a cube appearing, a 2-second pause, the cube playing an animation of the referent, then the participant proposing their input. The animation playing first removes the notion that the participant is directly interacting with the system. However, their belief that someone else is interacting with this system in a separate room, and the onscreen gesture aids (described later), caused the user to feel that this was a live system. 

The referents (i.e. actions) that were used included the canonical manipulations (i.e. selection, rotation, positioning) found in \cite{3duibook} and the interactions that would be commonly used in a 3D manipulation or building task. They include translation and rotation on each axis, scaling, selection, and the creation or deletion of an object. This study looks at the use case of a 3D environment such as an architecture application, where objects must be manipulated and placed within that environment. This can be extended into interactive learning environments or data visualization environments where manipulating virtual content can provide better learning outcomes \cite{PET+09}. Most optical see-through AR-HMDs (e.g., Magic Leap One) and some VR-HMDs (e.g., Oculus Quest) have built in ego-centric sensors. With that in mind, the gestures in this study were analyzed by viewing the ego-centric interactions within the environment.

The metrics used for gesture proposal interpretation are Agreement Rate (\AR)\footnote{Please note that agreement rate \AR uses a different font to avoid confusion with AR for augmented reality.}, co-agreement rate (\CAR) , and the (\VRD) significance test \cite{WOB+09, VAT+15, VAT+16}. \AR is the proportion of proposals in agreement over the total possible proposals pairs in agreement. High \AR can be interpreted as more consensus among participants in the proposals generated for a given referent. This metric is used at the referent level meaning that a given proposal will not have an associated \AR but a referent will. Based on distributions of \AR over various sample sizes participants an \AR of $0.3$ has been said to indicate high agreement given our N of $24$ \cite{VAT+15}. The \VRD is a test of the difference in agreement rates between $k$ referents. A low p-value indicates that there is a difference between the tested referents. The \CAR can be seen as the percent of participants that agree on proposals for $k$ referents. \textcolor{black}{Fleiss'} Kappa and the associated chance agreement term are used to justify using an \AR of $0.3$ as high \cite{TSA18}. 

For speech proposal analysis the consensus-distinct ratio (\CDR) and max-consensus (\MC) were used \cite{MOR12}. The \CDR is the percent of matching proposals that have been suggested by more than a recommended baseline of two participants out of all the proposals for a given referent \cite{MOR12}. \MC is equal to the percent of participants proposing the top-ranking proposal. The combination of these metrics can be used to see the peak and spread of speech proposals. 

\subsection{Participants}

The study consisted of 24 volunteers (10 Female, 14 Male). Participants were recruited using emails and word of mouth. Participants were 18 - 46 years old (Mean = 25, SD = 6.9). Six participants had less than half an hour of previous AR-HMD usage experience, the other participants had no prior device usage. All participants reported normal or corrected to normal vision. Five participants reported being left-handed. Five participants reported weekly use of VR. Only one of 2 of those participants used VR more than 5 hours weekly (5 hours, 10 hours), the rest were 1-3 hours weekly. 

\subsection{Procedure}

For each session participants started by completing the informed consent and demographic questionnaire. That questionnaire asked about prior device usage (AR, VR, multi-touch), age, handedness, vision, and gender. A two-minute instruction video was shown describing the experiment after which the participant could ask the experimenter questions. During the video, they were told that any utterance or gesture either one-handed or two-handed, produced was acceptable. The participant would then don the AR-HMD and complete a practice trial for each input modality. During the practice trials, the participant could ask any questions they had and adjust the device. Participants were also alerted to the devices gesture recognition aid shown (Figure \ref{fig:trialExample}) during the practice. This aid was an image of the outline of a left and right hand. The hands were white when a participant's corresponding hand was inside the device's gesture sensing range. They would be red with a line through them when the participant's corresponding hand was outside of the device's recognition range. This aid was provided to help prompt participants to generate gesture proposals that could be used in AR-HMDs as well as to add more immersion to the interactions with the object in the experiment. As this was a WoZ study, the aid was only adding realism to the task, no gestures were recognized.

The referents were shown as animations (showing the object then moving it left over 2 seconds for the referent \textit{move left}). \textcolor{black}{No text was shown to the user. For three referents animations that were not basic movements had to be shown. For the \textit{create} and \textit{delete} referents particle effects of an object appearing or disappearing over two seconds were used. For the \textit{select} referent, the object was highlighted by increasing its hue and adding a light outline.} Each referent was presented as a cube rendered 50cm in front of a user's display. The modality to use for the proposals was shown as text above the cube. The experimenter would trigger the loading of the next referent a few seconds after a proposal was generated by the participant. The new referent would always appear in the center of the participant's display, stay there for 2 seconds, then execute the animation for the referent.

\subsection{Apparatus}

This experiment was conducted using a Magic Leap One optical see-through AR-HMD. The WoZ system was developed in Unreal Engine 4.23.0. A Windows 10 professional computer with an Intel i9-9900k 3.6GHz processor and an Nvidia RTX 2080Ti graphics card was used for development.  Data were recorded on the Magic Leap One. A GoPro hero 7 black was used to record an ego-centric view of the interactions for analysis. A 4k camera was used to record an exo-centric view of the interactions as a backup to the GoPro.

\section{Results}


\subsection{Gestures Proposals}

\subsubsection{Gestures from the unimodal gesture block}

The average \AR observed for the gesture block was $0.302$ with $\kappa_F=.257$. Given our sample size of 24 and the low chance agreement term ($p_e =.052$) used in \textcolor{black}{Fleiss'} Kappa coefficient we consider rates above $0.3$ as high levels of consensus \cite{TSA18, VAT+15}. Agreement rates are shown in Table \ref{tbl:AR}.

\begin{table}[htbp]
\begin{center}
  \footnotesize
\caption{Agreement rates per referent by block}
\label{tbl:AR}
  \begin{tabular}{|p{2.18cm}|p{.3cm}|p{.3cm}|p{.3cm}|p{.3cm}|p{.3cm}|p{.3cm}|p{.3cm}|p{.3cm}|p{.3cm}|p{.3cm}|p{.3cm}|p{.3cm}|p{.3cm}|p{.3cm}|p{.3cm}|p{.3cm}|p{.3cm}|p{.3cm}|} 
 \hline
 & \rotatebox{90}{Create} & \rotatebox{90}{Delete} & \rotatebox{90}{Enlarge} & \rotatebox{90}{Move Away} & \rotatebox{90}{Move Down}
 & \rotatebox{90}{Move Left} & \rotatebox{90}{Move Right} & \rotatebox{90}{Move Towards  } &  \rotatebox{90}{Move Up} &  \rotatebox{90}{Pitch Down}
 &  \rotatebox{90}{Pitch Up} & \rotatebox{90}{Roll C} &\rotatebox{90}{Roll CC} &\rotatebox{90}{Select} & \rotatebox{90}{Shrink} & \rotatebox{90}{Yaw Left} & \rotatebox{90}{Yaw Right} \\
 \hline\hline
 Gesture & 0.21 & 0.11 & 0.28 & \cellcolor{blue!25}0.37 & \cellcolor{blue!25}0.38 & \cellcolor{blue!25}0.49 & \cellcolor{blue!25}0.44 & 0.28 & \cellcolor{blue!25}0.49 & 0.16 & 0.28 & \cellcolor{blue!25}0.56 & \cellcolor{blue!25}0.39 & 0.09 & 0.14 & 0.25 & 0.22\\
Gesture and Speech & 0.09 & 0.05 & 0.18 & \cellcolor{blue!25}0.50 & 0.29 & \cellcolor{blue!25}0.33 & \cellcolor{blue!25}0.32 & \cellcolor{blue!25}0.34 & \cellcolor{blue!25}0.35 & 0.19 & 0.22 & 0.28 & \cellcolor{blue!25}0.45 & 0.09 & 0.14 & 0.15 & 0.25\\
 \hline
\end{tabular}
\textbf{Legend}: C: clockwise, CC: counterclockwise, Highlighted cells have high agreement
\end{center}
\end{table}

\begin{figure}[htbp]
  \centering
  \includegraphics[width=\columnwidth]{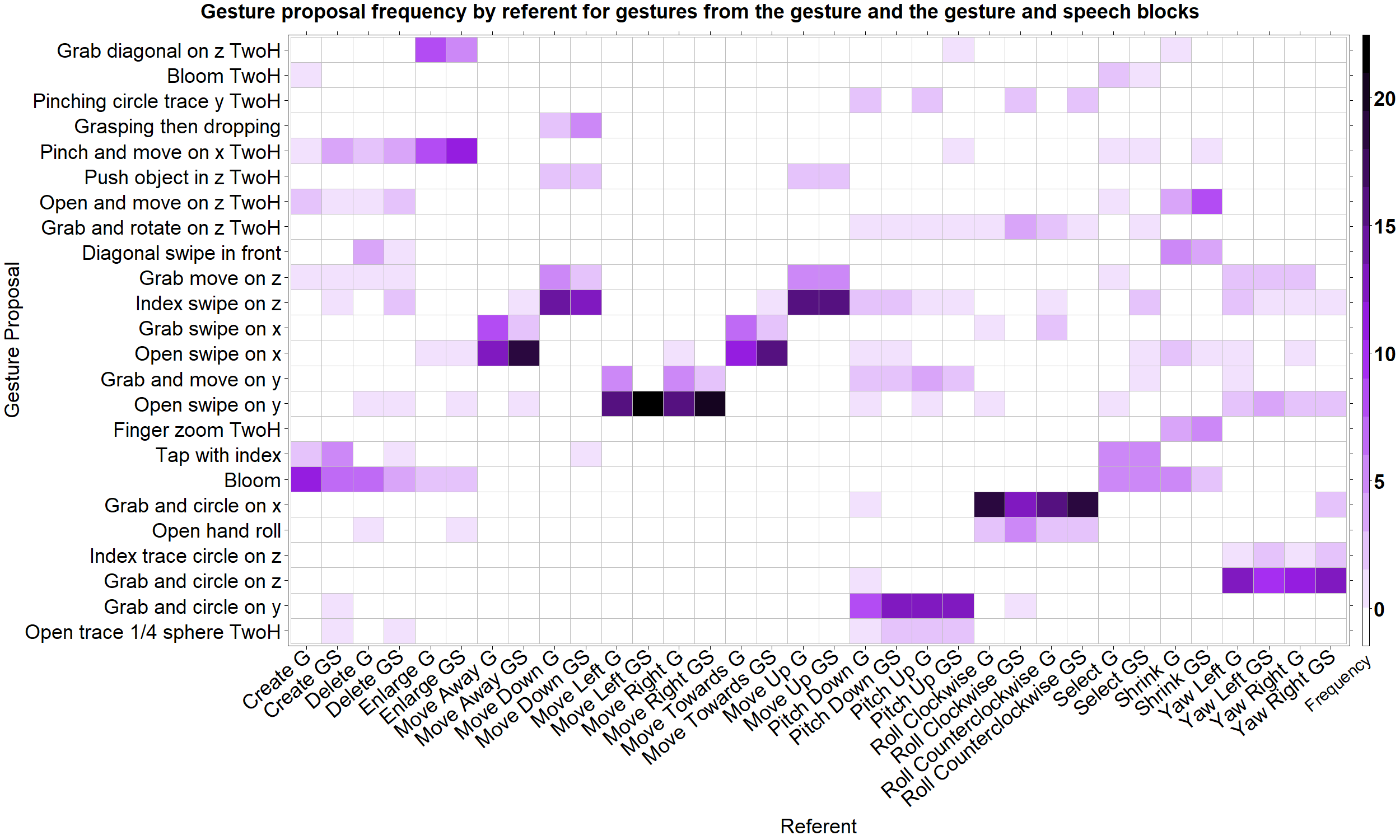}
  \caption{Gesture proposal frequency by referent for gestures from the gesture and the gesture and speech blocks}
  \label{fig:G_GS_Heatmap}
  \textbf{Legend}: G: Gesture Block, GS: Gesture and Speech block, TwoH: Two handed gesture, Open: fingers open, Grab: hand closed, Pinch: two or three finger pinching, z: up, x: forward, y: side 
  \Description{Agreement Rates by referent for gestures from the gesture and the gesture and speech blocks}
\end{figure}

The effect of referent type on agreement rates was observed to be significant ($V_{rd(16,N=408)}=510.342,p=.001$). High agreement was found for each of the translation referents except \textit{move away}, and for both the \textit{roll clockwise} and \textit{roll counterclockwise} referents (Table \ref{tbl:AR}). The highest \AR was found in the \textit{roll clockwise} referent ($\mathcal{AR}_{roll\:clockwise}=.56$). A mapping of the frequency of gesture proposals with more than three participants suggesting them and the corresponding referents can be seen in Figure \ref{fig:G_GS_Heatmap}. The gestures from the gesture block have ``G'' next to the referent name.

\begin{figure}[htbp]
  \centering
  \includegraphics[width=.9\columnwidth]{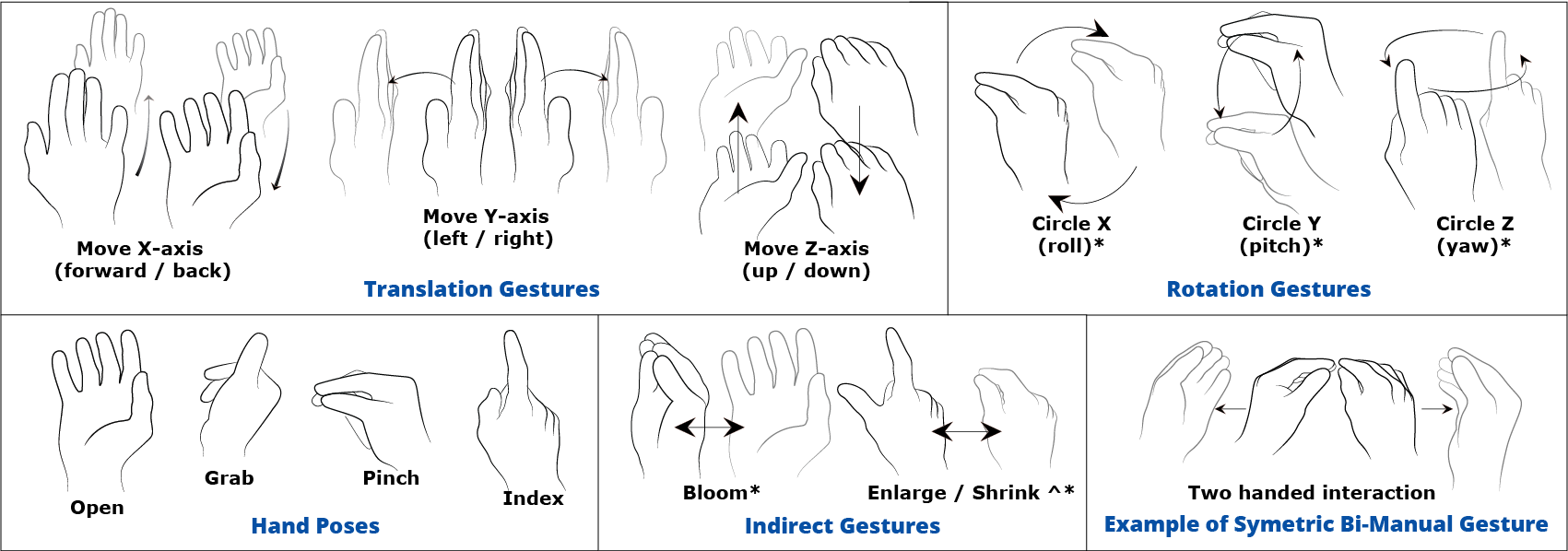}
  \caption{\textcolor{black}{Hand pose examples, two handed gesture example, and common gestures by category of movement or type of gesture}}
  \label{fig:Hands}
  \textcolor{black}{\textbf{Legend}: *: reversible gesture, $\wedge$: commonly two handed, z: up/down, x: forward/back, y: left/right
  \Description{Hand pose and movement examples by type of movement}}
\end{figure}

The more abstract referents, \textit{Create}, \textit{Delete}, and even \textit{select} exhibited low agreement rates ($\mathcal{AR}_{shrink}=.14$, $\mathcal{AR}_{delete}=.11$, $\mathcal{AR}_{Select}=.09$). This is mostly due to disagreement between proposals shown by an increase in the count of colored cells in Figure \ref{fig:G_GS_Heatmap}. Common hand poses and movements are shown in Figure \ref{fig:Hands}. \textit{Select} had low \AR due to participants having a difficult time interpreting the referent animation. \textit{select}'s animation showed the cube normally (left side of Figure~\ref{fig:trialExample}) then gradually becoming highlighted by reducing the hue after a 2-second delay. In pilot tests on the \textit{select} referent we attempted other visualizations such as bouncing, or an arrow appearing and pointing at the cube. These animations primed the speech and gesture produced. The highlight animation had the highest rate of participants guess what it was, but that rate was still fairly low. 

The translation referents (\textit{up, down, left, right, away}, and \textit{move away}) had high gesture agreement among participants ($\mathcal{AR}_{translations}=.432$). Among these translational referents, the direction of motion displayed a significant effect on agreement rates ($V_{rd(5,N=144)}=52.765,p<.001$). A significant difference in agreement was observed for referents \textit{towards} and \textit{away} ($V_{rd(1,N=48)}=9.921,p<.01$). \textit{Roll clockwise} and \textit{roll counterclockwise} had high \AR with an average ($\mathcal{AR}_{roll}=.475$). This was higher than the average \AR for all the rotational referents ($\mathcal{AR}_{rotations}=.31$) which drops to ($\mathcal{AR}_{rotations\:without\:roll}=.23$) when roll is removed. We believe that participants may not have had much experience with altering the pitch or yaw of virtual objects and this is reflected with the low \AR. The excepting being roll manipulations, which seem more common with objects like clock hands moving that way, inflating their \AR. 

There was low \AR for \textit{shrink} and \textit{expand}, which is surprising due to the prevalence of touchscreen phones and near-daily use of the two-finger zoom-in and zoom-out commands. Those gestures occurred with some frequency, however, there were a high number of two-handed comparable gestures proposed (Figure \ref{fig:G_GS_Heatmap}). For these people would pinch either corner and pull or push their hands away or towards other either diagonally or horizontally.  

The heatmap in Figure \ref{fig:G_GS_Heatmap} helps show the trends among gesture proposals, darker colors indicate more proposals. The gestures mapped are all reversible gestures meaning a movement in the opposite direction is the mirror of the gesture. An example of this is seen in the gesture for \textit{move up} which was a palm up push up where \textit{move down} was a palm down push down. The referents \textit{move left} and \textit{move right} had very few different proposals indicating high agreement on the appropriate gesture. Whereas, referents like \textit{select} had a high range of proposals given. When examining the plot horizontally by proposal instead of vertically by referent trends in how participants map the same gesture to multiple actions are seen. For example, an open hand swipe either left or right was used for 9 referents. The uses make sense, a quick swipe from right to left could be seen as deleting an object, or touching the side of an object and moving left or right would change its yaw. The ``Bloom'' gesture was used for every abstract referent. The variations present in some manipulations were only in the pose of the hand, or the number of hands, but not the motion of the gesture. \textit{Move up} had three common proposals with each centering around some sort of grab and a movement on the z-axis. 

\subsubsection{Gestures from the multimodal gestures and speech block}

The results for the gesture proposals from the gesture and speech are very similar to the gestures from the gesture alone block. By comparing columns with the matching referent names (e.g. \textit{create G} and \textit{create GS}), an image of the differences of proposals across these blocks can be drawn. The overall agreement rate observed for the gestures in the gesture and speech block was $0.247$ with $\kappa_F=.218$.  The low chance agreement term ($p_e =.037$) used in \textcolor{black}{Fleiss'} Kappa coefficient indicates an agreement beyond chance \cite{TSA18}, allowing us to consider \AR rates above $0.3$ as high \cite{VAT+15}. The agreement rates for each referent are shown in Table \ref{tbl:AR}.

The effect of referent type on agreement rates was observed to be significant ($V_{rd(16,N=408)}=904.091,p=.001$). High agreement was found for each of the translation referents except \textit{Move Down} $\mathcal{AR}_{Move\:Down}=.29$. This was caused by an increase in the number of ``drop'' gesture proposals. \textit{Roll counterclockwise} also exhibited high \AR ($\mathcal{AR}_{Roll\:counterclockwise}=.45$) (Figure \ref{tbl:AR}). The highest \AR was found in the \textit{Move Away} referent ($\mathcal{AR}_{Move\:Away}=.5$). A mapping of the frequency of the top gesture proposals and the corresponding referents can be seen in Figure \ref{fig:G_GS_Heatmap}.

The abstract referents \textit{Create}, \textit{Delete}, and \textit{select} exhibited low agreement rates ($\mathcal{AR}_{Create}=.09$, $\mathcal{AR}_{Delete}=.05$, $\mathcal{AR}_{Select}=.09$). This is mostly due to disagreement between proposals shown by an increase in the count of colored cells in Figure \ref{fig:G_GS_Heatmap}. As in the gesture block, \textit{select} had low \AR due to participants having difficulties interpreting the referent's animation. The translation referents (\textit{up, down, left, right, away}, and \textit{move away}) had high gesture agreement (average $\mathcal{AR}=.355$). A significant disparity was observed for referents \textit{roll clockwise} and \textit{roll counterclockwise} ($V_{rd(1,N=48)}=59.522,p=.001$). \textit{Roll clockwise} and \textit{roll counterclockwise} had high \AR with an average of ($\mathcal{AR}_{Roll}=.475$. This was higher than the average \AR for all the rotational referents ($\mathcal{AR}_{Rotations}=.31$) which drops to ($\mathcal{AR}_{Rotations\:without\:roll}=.23$) when roll is removed. We believe that participants may not have had much experience with altering the pitch or yaw of virtual objects and this is reflected with the low \AR. As in the gesture block the scale referents had low $\mathcal{AR}_{Shrink, Enlarge}=.18, .14$. 

The bulk of the gestures shown in Figure \ref{fig:G_GS_Heatmap} are direct manipulation gestures. Translations are concentrated in a few gestures where rotations are spread across more proposals. Even so, most rotation proposals involved tracing or moving a participant's hand in a circle. In the case of most of the referents, there was an increased spread of gesture proposals in the gesture and speech block. This was not the case for every referent, some such as \textit{move left} and \textit{roll counterclockwise} have a decreased number of proposals in the gesture and speech block. Largely the gestures used did not change drastically between the two blocks.

\subsection{Speech Proposals}

Displaying the referent in elicitation studies \cite{ORT+17} and reading the referent aloud in gesture and speech elicitation studies \cite{MOR12} both have precedence. These practices can prime the utterances proposed. When interpreting these results remember that neither think out-loud protocol nor text was used for referents. The participant only saw an animation of the referent being executed. When analyzing speech proposals we have dropped the object specifier to remove a level of increased proposal complexity. We believe that if an object is already selected, using the command "Move the cube right" and "move right" could be reasonably considered the same, the exception being the \textit{select} referent. 

\subsubsection{Speech from the unimodal speech block} 

\begin{table}[htbp]
\begin{center}
  \footnotesize
\caption{Frequency of syntax format by block}
\label{tbl:syntaxRates}
  \begin{tabular}{|c | c | c | c | c | c | c |} 
 \hline
 & $<$action$>$ & \vtop{\hbox{\strut $<$action$>$}\hbox{\strut  $<$direction$>$ }}& \vtop{\hbox{\strut $<$action$>$ $<$object$>$ }\hbox{\strut $<$direction$>$}}  &  \vtop{\hbox{\strut $<$action$>$}\hbox{\strut  $<$object$>$ }} &  $<$direction$>$\\
 \hline\hline
 Speech & 28.19\% & 47.06\% & 14.22\% & 9.31\% & 1.23\% \\ 
 Gesture and speech & 38.48\% & 39.95\% & 12.99\% & 6.86\% & 1.72\% \\
 \hline
\end{tabular}
\end{center}
\end{table}

While were told that any utterance or sentence was acceptable, they primarily stuck to $<$action$>$ $<$direction$>$ or $<$action$>$ $<$direction$>$ syntax structure. The rates for syntax are found in table \ref{tbl:syntaxRates}. The difference between $<$action$>$ $<$direction$>$ and $<$action$>$ $<$object$>$ $<$direction$>$ was only a descriptive specifier of the object (e.g. ``cube''). The  $<$action$>$ and $<$direction$>$ words were the same as found when no specifier was used (e.g. ``move the cube left'' would be ``move left'').

The \MC and \CDR for this block are shown in Figure \ref{tbl:GS_S_proposals}. Note that \MC is equal to the percentage of participants proposing the top proposal per referent, shown in the "Top proposal" column in Table \ref{tbl:GS_S_proposals}. \textit{Yaw} referents had some of the highest \CDR indicating high disagreement among participants on the utterances proposed \textcolor{black}{($\mathit{\mathcal{CDR}_{Yaw\:left, Yaw\:right}=.62, .78}$)}. \textit{Delete} also had a high amount of disagreement among proposals ($\mathcal{CDR}_{Delete}=.57$).  Both \textit{create} and \textit{shrink} had low \CDR ($\mathcal{CDR}_{Create,\:Shrink}=.18, .25$). Low \CDR means that most participants grouped around the top proposals. The rest of the referents hold moderate disagreement values.

The highest \MC value belongs to \textit{move up} ($\mathcal{MC}_{move\:up}=.54$). Most participants proposed either "Move up" (54.17\%) or "go up" (12.5\%). The full list of each referent's top two proposals and the percent of participants proposing them can be seen in Table \ref{tbl:GS_S_proposals}. For the translational referents "move" was used as the $<$action$>$ command in either the top or second place proposal. \textit{Move down} ($\mathcal{MC}_{move\:down}=33.33\%$), which had ``drop'' as the top proposal, was the only translational referent that did not have ``move'' in it. The second-place proposal for \textit{move down} was ``move down'' ($29.17\%$ proposed). The referents for \textit{move up}, \textit{left}, and \textit{right} all had the directional term (up, left, right, down) included. \textit{Move towards} and \textit{move away} had either towards, and forward, or away, and back proposed as the $<$direction$>$ term. This indicates that aliasing ``away'' with ``back'', and ``towards'' with ``forward''. Aliasing commands has been suggested as being beneficial when dealing with unimodal speech \cite{WOB+09, MOR12}. Note that these terms are reversible, which was a common trend with most opposite proposals (e.g. ``appear'', ``disappear''). 


\begin{table*}[htbp]
\footnotesize
\caption{\textcolor{black}{Speech proposals for the speech from the speech block and the speech from the gesture and speech block}}
\label{tbl:GS_S_proposals}
\begin{tabular}{|p{1.37cm}||p{1.18cm}|p{.6cm}|p{1.18cm}|p{.6cm}|p{.6cm}||p{1.18cm}|p{.6cm}|p{1.2cm}|p{.6cm}|p{.6cm}|} \hline
& \multicolumn{5}{c||}{Speech from the speech block} & \multicolumn{5}{c|}{Speech from the gesture and speech block} \\\hline      
Referent     & Top proposal & \MC & 2nd place & \MC & \CDR & Top proposal & \MC & 2nd place  & \MC & \CDR \\ \hline\hline
Create & appear & 41.67\% & create & 20.83\% & 0.18 & appear & 33.33\% & create & 29.17\% & 0.18 \\ \hline
Delete & disappear & 50\% & remove & 16.67\% & 0.57 & disappear & 54.17\% & make disappear & 12.5\% & 0.33 \\ \hline
Enlarge & enlarge & 37.5\% & grow & 16.67\% & 0.36 & enlarge & 25\% & grow & 20.83\% & 0.56 \\ \hline
Move Away & move back & 25\% & move away & 12.5\% & 0.38 & move back & 16.67\% & push away & 16.67\% & 0.64 \\ \hline
Move Down & drop & 33.33\% & move down & 29.17\% & 0.44 & drop & 29.17\% & move down & 16.67\% & 0.46 \\ \hline
Move Left & move left & 37.5\% & slide left & 20.83\% & 0.44 & move left & 25\% & slide left & 16.67\% & 0.2 \\ \hline
Move Right & move right & 41.67\% & slide right & 20.83\% & 0.44 & move right & 20.83\% & slide right & 20.83\% & 0.33 \\ \hline
Move Towards & move forward & 20.83\% & move towards & 12.5\% & 0.36 & move forward & 16.67\% & move towards & 12.5\% & 0.43 \\ \hline
Move Up & move up & 54.17\% & go up & 12.5\% & 0.33 & move up & 41.67\% & go up & 8.33\% & 0.33 \\ \hline
Pitch Down & rotate & 20.83\% & rotate towards & 16.67\% & 0.46 & spin forward & 20.83\% & rotate towards & 16.67\% & 0.6 \\ \hline
Pitch Up & rotate away & 16.67\% & spin backward & 12.5\% & 0.5 & spin back & 16.67\% & rotate & 12.5\% & 0.43 \\ \hline
Roll C & spin right & 20.83\% & rotate & 16.67\% & 0.5 & rotate & 20.83\% & rotate right & 16.67\% & 0.36 \\ \hline
Roll CC & spin left & 25\% & rotate left & 20.83\% & 0.4 & spin left & 25\% & rotate & 16.67\% & 0.23 \\ \hline
Select & glow & 20.83\% & highlight & 20.83\% & 0.55 & change & 25\% & glow & 25\% & 0.36 \\  \hline
Shrink & shrink & 45.83\% & minimize & 8.33\% & 0.25 & shrink & 41.67\% & make smaller & 8.33\% & 0.23 \\ \hline
Yaw Left & spin left & 33.33\% & rotate & 16.67\% & 0.62 & spin & 29.17\% & rotate left & 16.67\% & 0.36 \\ \hline
Yaw Right & spin right & 29.17\% & rotate & 12.5\% & 0.78 & rotate right & 20.83\% & spin & 20.83\% & 0.6 \\ \hline          
\end{tabular}
\textcolor{black}{\textbf{Legend}: C: Clockwise, CC: Counterclockwise, $\mathit{\mathcal{MC}}$: Max-Consensus, $\mathit{\mathcal{CDR}}$: Consensus-Distinct Ratio}
\end{table*}

For the rotational referents (\textit{pitch}, \textit{roll}, \textit{yaw}) the average \MC was $24.31\%$ which is lower than the translations average \MC of $35.42$. For each rotation the action was specified by either ``spin'' or ``rotate'' in all of the top proposals by participants (Table \ref{tbl:GS_S_proposals}). This is not unexpected, the terms ``roll'', ``pitch'', and ``yaw'' are uncommon in most fields. \textit{Pitch} has the most unique mapping of proposals commonly ``towards'', ``away'' for \textit{pitch up} and ``back'' for \textit{pitch down}. \textit{Roll} and \textit{yaw} have the terms ``left'' and ``right'' for directions. We believe that this ambiguity is solved by adding gestures to indicate the ``spin'' direction, or by an expert assigning speech commands such as ``spin clockwise'' in the \textit{roll clockwise}. 

The referents \textit{create} and \textit{delete} had single word commands for the top and second place proposals as well as some of the higher \MC found ($\mathcal{MC}_{create, delete}=41.67\%, 50\%$). The top proposals were ``appear'' and ``disppear''. These proposals could be considered similar to the reversible gestures found in this study and others \cite{WOB+09, PIU+13}. ``Create'' appeared as a second place proposal $(20.83\%)$ and ``delete'' was a third place proposal $(12.5\%)$. \textit{Shrink}  ($\mathcal{MC}_{shrink}=45.83\%$) also had a high agreement between participants. As did \textit{enlarge} ($\mathcal{MC}_{enlarge}=37.5\%$). \textit{Select}, with its difficulties in animating had low agreement and high spread of proposals  ($\mathcal{CDR, MC}_{select}=.55, 21\%$).

\subsubsection{Speech from the multimodal gesture and speech block}

A chi-square test of independence showed that there was a significant association between the block and syntax choice $(X^2 (4, N=408)=10.928, p<0.03)$. Participants used a higher rate of $<$action$>$ only syntax than found in unimodal speech. $<$Action$>$ $<$direction$>$ syntax use was reduced by $7.11\%$. The rates for the syntax are found in Table \ref{tbl:syntaxRates}. Both of the syntax structures that used an $<$object$>$ specifier were lower in this block. Most often when an object would have been specified it was replaced by a gesture indicating the object. This gesture was often reaching out and grabbing or another type of direct manipulation.

The average \MC for the translational referents decreased by $10.33\%$ from the speech block (Figure \ref{tbl:GS_S_proposals}). This was due to more participants using the $<$action$>$ syntax. The \CDR did increase in the translational referents as well. Participants had less agreement on the appropriate proposal and the spread of proposals was wider.  Even with the differences in syntax use between blocks, the top choice proposals remained the same.

The rotational average \MC only decreased by $2\%$, the \CDR decreased by $0.113$. This means that while agreement on the top choice proposal was negligibly impacted between blocks, the spread of proposals given in the gesture and speech block for rotations was narrower than in the speech block. Most of the top choice proposals for translations changed between the two blocks (Table \ref{tbl:GS_S_proposals}). Some switched from using ``spin'' to ``rotate'' or vice versa. As an example, the proposal for \textit{yaw right} switched from ``spin'' to ``rotate'' while the top proposal for \textit{roll clockwise} did the opposite. We take this to mean that the words ``rotate'' and ``spin'' are without a clear mapping in participants' minds. For translations gesturing removes much of the ambiguity by allowing for a physical motion to indicate the intended rotation direction. 

Most proposals remained the same between the two blocks with slightly different \MC rates. There was a shift in \textit{create} from the top choice proposal of ``appear'' from ($\mathcal{MC}_{Create}=41.67$) to ($\mathcal{MC}_{create}=33.33$) in the gesture and speech block. This is captured in an increase of $8.34\%$ in the second choice proposal in the gesture and speech block. \textit{Delete} was mostly unchanged in top proposals but did have a decreased \CDR ($\mathcal{CDR}_{Delete}=.33$). Meaning there were less distinct proposals made. \textit{Enlarge} had a lower \MC and higher \CDR in the gesture and speech block ($\mathcal{MC, CDR}_{enlarge}=37.5\%, .56$).

\subsection{Co-occurring gestures and speech proposals}

When looking at pairings of speech and gesture proposals in the gesture and speech block the agreement rates fall drastically due to the highly nuanced nature of speech. Individually each modality had referents that experienced high levels of agreement. For gestures refer to Figure \ref{fig:G_GS_Heatmap} and Table \ref{tbl:AR}. For speech consensus refer to Table \ref{tbl:GS_S_proposals}. We feel that matching common syntax structure with gestures when looking at multimodal gesture and speech interactions is more beneficial than observing the pairing of utterances with gesture proposals. The speech syntax by block is shown in Table \ref{tbl:syntaxRates}. Gesturing remains consistent in both conditions indicated by a high p-value in a chi-square test $(X^2 (49, N=408)=10.928, p<0.247)$ (Comparing G and GS in Figure \ref{fig:G_GS_Heatmap}). The same is true of speech (Compare S and GS in Table\ref{tbl:GS_S_proposals}). This is beneficial in a few ways. In the case of translations and scaling it allows each input to serve as a back up to the other. Allowing for mutual disambiguation as found by \cite{OVI00}. In the case of rotations, the gesture provides context on the direction of rotation while the speech was commonly ``spin'' and a direction. With abstract commands, the same gesture, a ``bloom'' gesture, was found for multiple referents. In those cases, speech allows interpretation of which command is being executed with the gesture. 

\subsubsection{Timing of co-occurring gestures and speech}

In the gesture and speech block the time windows of phases of a co-occurring gesture and speech interaction were measured based on the time of gesture initiation. These were collected from videos of the experiment and hand-annotated. The phases used to describe interactions are gesture initiation, stroke start, speech start, stroke stop, and speech stop. These are taken from McNeil's segmentation of co-occurring gesture and speech interactions \cite{MCN05}. The gesture start is the first perceptible movement made by someone. Speech start is the first perceptible sound being made. For both of those if a false start was found it was discarded and the time of the next movement was taken. As an example, if a participant said ``Ummm'' then later said ``move'', the time of ``move'' was used. A stroke is considered to be the segment of a gesture that holds the information content of the gesture, as well as the peak of effort in that gesture \cite{MCN05}. Gesture stroke was found by measuring the time of the first visible change in the direction of the gesture. The stroke stop was the last change in direction and was found by reversing from the end of a gesture. A full gesture interaction would look like someone starting to move their hand in preparation for a stroke (gesture start), starting a meaningful gesture (stroke start), then ending the gesture (stroke stop). The hand moves up in preparation, pushing the object forward, then retracts to its initial state.

\begin{figure}[htbp]
  \centering
  \footnotesize
  \includegraphics[width=.7\columnwidth]{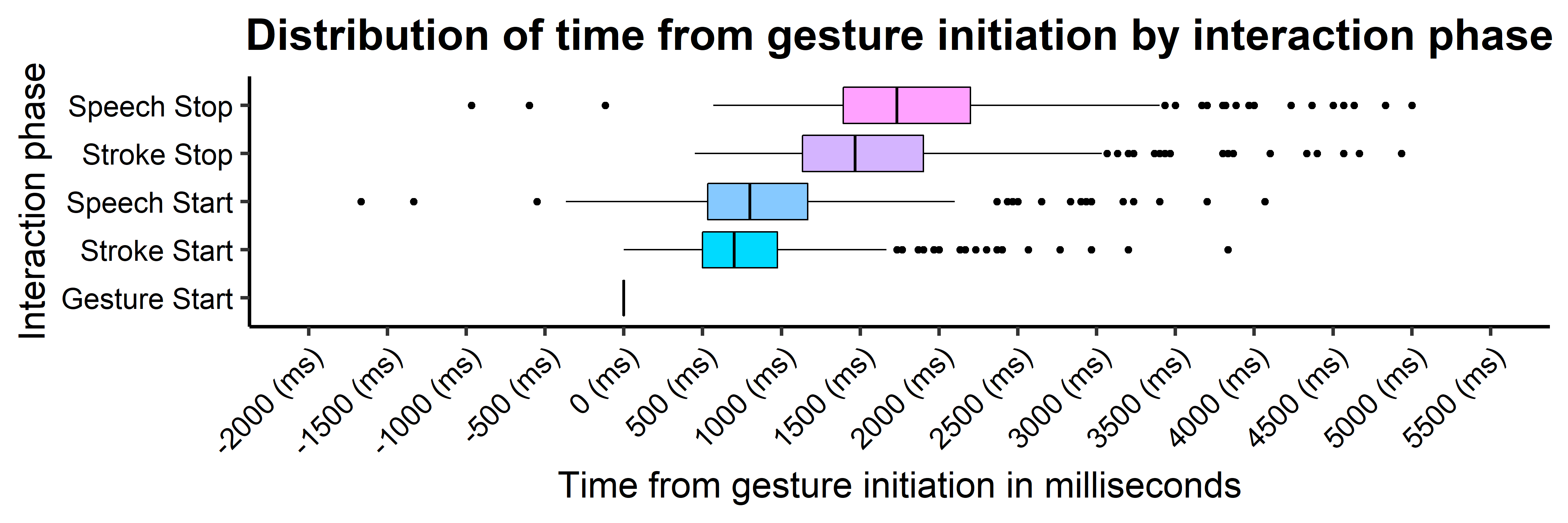}
  \caption{Distribution of time from gesture initiation by interaction phase}
  \label{fig:Boxplot_interactionTimes}
  \Description{Distribution of time from gesture initiation by interaction phase}
\end{figure}

Shapiro-Wilks tests show that the time information took a non-normal distribution of each of the phases at $(p < .001)$. Bonferroni adjusted Wilcoxon rank-sum pairwise comparisons indicate that each phase's time is significantly different from each other. The p-values were $(p < 0.001)$ in each comparison except between ``stroke start'' and ``speech start'' which was $(p = 0.03)$. The descriptive statistics for times by the phase of interaction are shown in Table \ref{tbl:interactionTimes}. 

We find that in this experiment speech nearly always occurs after a gesture is started (Figure \ref{fig:Boxplot_interactionTimes}). The difference in start time is around $816.67$ \textcolor{black}{ms}. Importantly, the information content of the gesture, the stroke, starts commonly $908.72$ \textcolor{black}{ms} after gesture start (Table \ref{tbl:interactionTimes}). This means that by watching a gesture's changes in direction, we can predict when speech will occur, and when a meaningful message is communicated. Strokes were found to end before speech $237.39$ \textcolor{black}{ms}. The total interaction from start to finish was typically $1875.66$ \textcolor{black}{ms}. Most speech proposals were only $2$ words so this relatively short interaction time makes sense.

\begin{table}[htbp]
\begin{center}
  \footnotesize
\caption{Time from gesture start for phases of an interaction in milliseconds}
\label{tbl:interactionTimes}
 \begin{tabular}{|c c c c c c|} 
 \hline
 & Gesture Start & Stroke Start & Speech Start & Stroke Stop & Speech Stop \\
 \hline\hline
 Mean & 0 & 908.72 & 816.67 & 1638.27 & 1875.66 \\
 Standard Deviation & 0 & 705.48 & 547.43 & 787.42 & 800.64 \\
 Standard Error & 0 & 34.93 & 27.10 & 38.98 & 39.64 \\
\hline 

\end{tabular}
\end{center}
\end{table}

These results are similar to previous work \cite{BOU+98, LEE+08}, though slightly quicker and more granular. These results expand time windows from being formed for pointing gestures only \cite{BOU+98}, and show that these time windows follow similar patterns for deictic and manipulative gestures. They also show that gesture and speech interactions in AR-HMDS have similar timings \cite{LOE12} and patterns of occurrence \cite{SCH84} as in other environments.

\section{Discussion}

\textcolor{black}{The hand positions found here were similar to the ones observed by Piumsomboon et al. \cite{PIU+13}. The gesture proposals were commonly single-handed. This is similar to findings on multi-touch surfaces \cite{MOS+08, KIN+09, MOR+10} and mid-air full-body studies \cite{ORT+17}. For manipulations users often interacted based on that actions real-world corollary. This is evident in the translation gestures which were predominantly some form of directly pushing the surface of the object. This theme of interaction was seen with manipulation gestures in previous work \cite{PIU+13}. We speculate that the similarities in proposals were due to the object being rendered in the participant's real-world view by use of optical see-through AR. With that, users would interact based on their interpretations of naïve physics when possible \cite{JAC+08}. This was mostly true for rotations which were accomplished by either grabbing some part of the object and moving their hand in circle motions as also seen in Piumsomboon et al.'s study \cite{PIU+13}. The exception to these similarities is in the occurrence of ``index extended'' circular motions as an indirect gesture.}

\textcolor{black}{Scaling was often a two-handed pinch and drag gesture which was more common than touch screen ``zoom in'' and ``zoom out'' gestures. Grabbing the corners or sides of an object would correspond with how a mental model of a stretchable object would be manipulated. This gesture was seen for scaling on an axis in \cite{PIU+13}. Similarities in gesture proposals between these studies start disappearing as the referents become more abstract. This can be seen when comparing the proposals for \textit{delete} which was a ``grasping'' gesture in other work \cite{PIU+13} and a ``bloom'' gesture here.}

\textcolor{black}{That most of these gesture proposals extend across two studies and two-time points is a strong indication that these gestures and hand poses should be candidates for inclusion in future AR interaction systems. This study did not ask participants to reserve proposals for a single interaction (i.e., a bloom could be used for \textit{create} and for \textit{select}). Redundantly mapped proposals showed up more in the abstract referents. In the work of Piumsomboon et al. participants were asked to refrain from redundantly mapping inputs \cite{PIU+13}. The similarities of proposals between these works show that requiring unique interactions may not have greatly impacted many of the gesture proposals \cite{PIU+13}. An interesting, redundantly mapped gesture was the ``index swipe'' which was used for both \textit{yaw} and \textit{move up/down}.}

We feel that the combination of high levels of agreement for translations in the gesture block and the tendency to have more unique proposals given in the gesture and speech block indicate that unimodal gestures are well suited for object manipulations. While rotations had a high number of single-hand ``grab and rotate'' gestures, many were indirect manipulations using a index finger and tracing a circle. For these, a non-isomorphic gesture seem well suited. The most agreed-upon proposals for manipulations were all reversible gestures. Indicating a preference for reversible gestures which mirrors previous work \cite{WOB+09, PIU+13}.

\textcolor{black}{
Some of these direct manipulations were implemented and tested against a gesture+speech interface in the work of Piumsomboon et al. \cite{PIU+14}. The findings were similar to the user stated expectations observed here. When specific degrees or units were needed participants indicated a preference for speech. For most basic or single object manipulations, gesture seemed preferred across both studies \cite{PIU+14}. Peoples' preference for multimodal interactions typically increases as a task's cognitive load increases \cite{OVI+04} or the task's complexity increases \cite{PIU+14}. We expect that if more complex referents were used the user stated preference for multimodal interactions would have been higher.
}

Gestures showed less usability for the \textit{create} and \textit{delete} referents. Speech had more clarity in these cases with common utterance being ``appear'' and ``disappear''. Gesture proposals for abstract referents were consistently the ``bloom'' gesture, which was proposed for many referents, and thus hard to interpret without additional context. Speech show more promise for use with abstract commands and conceptually difficult actions that do not map well to a user's mental model. An example would be opening a new browser window, which was not tested here. Speech proposals for both \textit{create} and \textit{delete} had high agreement, emphasizing this strength.


When used together gestures and speech provide different benefits based on the type of referent being executed. For translations and scaling this was commonly redundancy, which allows for error correction in a recognizer system. For rotations, this pairing allows a clear communication of the desire to rotate then clarifying the direction with a co-occurring gesture. This allows for intuitive interactions with mutual disambiguation with information from the complementary channel. An added benefit of allowing speech and gesture for rotations is the ability for participants to communicate the degrees of rotation, allowing for more accurate interactions. 

\textcolor{black}{In the speech condition participants preferred to use $<$action$>$ $<$direction$>$ or $<$action$>$ $<$object$>$ $<$direction$>$ syntax over complete sentences. Implying that both unimodal and multimodal speech utterances are syntactically simplified compared to conversational speech \cite{OVI99}. This is seen as saying ``move'' and ``finger flicking'' in the direction of the intended movement. In either case, the intended  $<$action$>$ was present indicating that full natural language processing may not be necessary for basic multimodal interactions.}


\textcolor{black}{
This work contributes to findings on multimodal interactions and touches on some of the potential pitfalls of referent display which would cause reproduction to be difficult, as mentioned by Villarreal-Narvaez et al. \cite{VIL+20}. The impact of referent display on proposals is seen most saliently in the low \AR for the \textit{select} referent which often received high \AR in prior studies \cite{PIU+13, ORT+19}. The timing information and patterns here provide insight into the formation of these interactions and extends the timing windows constructed by Lee et al. \cite{LEE+08} by adding the phase of the interaction by the time of that phases initiation. This study gathers proposals within each modality allowing for comparison against gesture only studies \cite{PIU+13}, while also contributing to the less common multimodal elicitation literature \cite{KHA+19, MOR12}. 
}

\section{Design Guidelines}

Instead of directly proposing a single set of consensus interactions within each modality we have chosen to show the distribution of interactions. By looking at these distributions a picture of trends across the top few proposals can be seen. For some referents, such as the translational referents, the top gesture in the gesture and the gesture and speech block matched (Figure \ref{fig:G_GS_Heatmap}). For translations often the top proposal was a reversible swiping gesture for moving the object in the x-axis and y-axis and an index extended swipe for movement on the z-axis. The speech proposals in these cases were also reversible (Figure \ref{tbl:GS_S_proposals}). The first choice in all translations except \textit{move down} was to say ``move'' and then a direction. For \textit{move down} people commonly said ``drop''. \textit{Create} and \textit{destroy} followed the same pattern with a reversible bloom gesture either starting closed then opening or starting open then closing and the utterances ``appear'', and ``disappear''. For most gestures, a bi-manual version that was a symmetric two-handed version of the uni-manual proposal was also proposed (i.e. pushing with one open hand versus pushing with two). 

\textcolor{black}{
Most gestures were based on the participants' understanding of naive physics, meaning how they perceived an object would react to an interaction as it would in the real world. Most variations occurred within specific hand poses but not the larger movements of the hand/arm. As such we recommend aliasing manipulative gestures across hand positions (open hand, pinch, grab) based on the type of movement. A second consideration should be made on the inclusion of bi-manual gestures. while not found in abundance here, other work \cite{KHA+19, PIU+13} has found evidence that users may gravitate towards using them in other domains and with larger objects \cite{TAR+18, PLA+17}.
}


Other referents had less consistency.  In the case of \textit{shrink} and \textit{enlarge} a ``bloom'' gesture and two handed ``pinch and drag'' gestures were common. In this case, we would suggest reserving the ``bloom'' gesture for \textit{create / delete} and allowing ``grab and pull'' and scaling as seen both here and in earlier work \cite{PIU+13}. The top speech proposals for scaling were more agreed upon and should be implemented as well. Those were the reversible pair ``enlarge'', and ``shrink''. Rotational referents other than \textit{roll clockwise} have high levels of disagreement among proposals. ``spin'' and ``flip'' should be enabled as action selection words then a gesture should be allowed for controlling the direction of the rotation. 

\textcolor{black}{Direct manipulations should be allowed when possible, especially for basic manipulations. Speech and gesture as multimodal interactions showed promise in areas where one or the other input lacked and should be included. Implementing a system such that it has an internal model of functionality that aligns with what most participants formed as their mental model of functionality would increase the user's chances of guessing the inputs. This would be most easily achieved with direct manipulations, which in this study were often very close to their real-world corollary.}

Participants seldom used full sentences or referred to the object being manipulated (Table \ref{tbl:syntaxRates}). Due to that word spotting should be sufficient for most tasks. Only two participants used full sentences and those sentences followed the  $<$action$>$ $<$object$>$ $<$direction$>$ syntax with prepositional terms added (e.g. ``move to the right'' compared to ``move right''). In either command, the actual information content is held in the $<$action$>$ $<$direction$>$ terms which could be spotted. The use of simple commands when possible was also observed by \cite{LEE+08}.

The windows built around co-occurring interactions are incredibly useful to systems needing to decipher interactions. With segmenting interactions based on the first movement of a gesture, the transition into the stroke phase, the information content of both the speech and the gesture portions of the interaction can be found. In this study gestures nearly always preceded speech (405/408 proposals). Most commonly speech was around $816.7$ milliseconds after a gesture initiated. The stroke was often $908.7$ milliseconds after the start of a gesture. Both of those phases represent the initiation of the actual information content of the interaction. The back end of these interactions is slightly less concrete. Often the end of a gesture preceded the end of an utterance. A system could be designed to use a time-out window after which the speech would be considered a separate interaction.

\section{Limitations of the Study}

By choosing to show animations for referents the gesture proposed may be biased to follow the animations shown. This choice was made to preserve the value of the speech proposals with pilot studies that showed speech was less impacted when showing the animations of the referents as opposed to the text. This study only allowed one proposal per referent per block. Having participants propose more than one interaction may have generated interactions that they felt more well suited to the referents. This study only showed a single virtual object at a time, which would impact the selection phase of any interaction. To help compensate for this we used the referent \textit{select} independently. 

\textcolor{black}{
For the rotational referents participants would sometimes use misaligned gestures and speech. They might say ``roll clockwise'' and perform a counterclockwise movement with their hand. Multimodal systems can suffer from compounding errors caused by incorrect recognition, or mismatched interactions such as the ones seen in this study \cite{BOU06}. These errors could take more time than standard uni-modal errors to correct or cause compounding errors when a second error is made during an attempt to correct the first. 
}


\section{Conclusions and Future Work}

Several questions remain unanswered. If there were more than one object shown the gesture results would show more selection gestures. The choice of an object used could also impact the production of interactions. If a larger object or a differently shaped object was presented the hand postures used may differ. Future work should involve testing the proposals found here against ones produced by text-based referents to assess the impact of referent display. 

\textcolor{black}{
Compound errors in uni-modal text entry systems cause a generally linear increase in correction time \cite{SAB+10}. Recent work has shown that improved error correction methods can reduce the time it takes users to reconcile text entry errors, decreasing the overall amount the user is slowed down by the error correction process \cite{OHU+19}. Further work is needed to examine whether this holds true for multimodal interactions.
}

This work presents a within-subjects elicitation study across three input modalities (gestures, speech, and co-occurring gesture and speech). By examining each modality independently direct comparisons between the changes in speech and gesture from unimodal interactions to multimodal interactions are shown. Trends in gesture proposals are shown at a granular level. Highlighting that while there is often disagreement in proposals given, that disagreement manifests as variations in with similar underlying formations. In gestures, this was a variation of the hand position and not in the gross movement. In speech, this disagreement is seen as consistency in the $<$direction$>$ phrases used and minor variations in the $<$action$>$ phrase (e.g. ``move'' to ``go''). While a singular mapping of the top proposals would yield a consensus set that is discoverable to most users, by aliasing and understanding the likely variations in interactions, a larger percentage of users' natural interaction preferences can be captured. 

This work extends the work of linguists \cite{MCN05, KEL+10, MOL+12}, and the work of computer scientists \cite{BOL80, COR+05, BOU+98, LEE+13} into AR-HMD building environments. Timing windows based on the phases of co-occurring gesture and speech interactions as described by McNeil \cite{MCN05} have been constructed. Showing that in HCI the gesture stroke is closely aligned with the information content of both the gesture and the utterance given. These windows can be used to construct more accurate multimodal fusion models.

\begin{acks}
This work was supported by the National Science Foundation (NSF) awards NSF IIS-1948254, and NSF BCS-1928502 and the Defense Advanced Research Projects Agency (DARPA) ARO contract W911NF-15-1-0459.
\end{acks}

\bibliographystyle{ACM-Reference-Format}
\bibliography{manuscript}


\begin{thebibliography}{63}


\ifx \showCODEN    \undefined \def \showCODEN     #1{\unskip}     \fi
\ifx \showDOI      \undefined \def \showDOI       #1{#1}\fi
\ifx \showISBNx    \undefined \def \showISBNx     #1{\unskip}     \fi
\ifx \showISBNxiii \undefined \def \showISBNxiii  #1{\unskip}     \fi
\ifx \showISSN     \undefined \def \showISSN      #1{\unskip}     \fi
\ifx \showLCCN     \undefined \def \showLCCN      #1{\unskip}     \fi
\ifx \shownote     \undefined \def \shownote      #1{#1}          \fi
\ifx \showarticletitle \undefined \def \showarticletitle #1{#1}   \fi
\ifx \showURL      \undefined \def \showURL       {\relax}        \fi
\providecommand\bibfield[2]{#2}
\providecommand\bibinfo[2]{#2}
\providecommand\natexlab[1]{#1}
\providecommand\showeprint[2][]{arXiv:#2}

\bibitem[\protect\citeauthoryear{Alharbi, Arif, Stuerzlinger, Dunlop, and
  Komninos}{Alharbi et~al\mbox{.}}{2019}]%
        {OHU+19}
\bibfield{author}{\bibinfo{person}{Ohoud Alharbi},
  \bibinfo{person}{Ahmed~Sabbir Arif}, \bibinfo{person}{Wolfgang Stuerzlinger},
  \bibinfo{person}{Mark~D. Dunlop}, {and} \bibinfo{person}{Andreas Komninos}.}
  \bibinfo{year}{2019}\natexlab{}.
\newblock \showarticletitle{WiseType: A Tablet Keyboard with Color-Coded
  Visualization and Various Editing Options for Error Correction}. In
  \bibinfo{booktitle}{\emph{Proceedings of the 45th Graphics Interface
  Conference on Proceedings of Graphics Interface 2019}} (Kingston, Canada)
  \emph{(\bibinfo{series}{GI'19})}. \bibinfo{publisher}{Canadian Human-Computer
  Communications Society}, \bibinfo{address}{Waterloo, CAN}, Article
  \bibinfo{articleno}{4}, \bibinfo{numpages}{10}~pages.
\newblock
\showISBNx{9780994786845}
\urldef\tempurl%
\url{https://doi.org/10.20380/GI2019.04}
\showDOI{\tempurl}


\bibitem[\protect\citeauthoryear{Anastasiou, Jian, and Zhekova}{Anastasiou
  et~al\mbox{.}}{2012}]%
        {ANA+12}
\bibfield{author}{\bibinfo{person}{Dimitra Anastasiou}, \bibinfo{person}{Cui
  Jian}, {and} \bibinfo{person}{Desislava Zhekova}.}
  \bibinfo{year}{2012}\natexlab{}.
\newblock \showarticletitle{Speech and Gesture Interaction in an Ambient
  Assisted Living Lab}. In \bibinfo{booktitle}{\emph{Proceedings of the 1st
  Workshop on Speech and Multimodal Interaction in Assistive Environments}}
  (Jeju, Republic of Korea) \emph{(\bibinfo{series}{SMIAE '12})}.
  \bibinfo{publisher}{Association for Computational Linguistics},
  \bibinfo{address}{Stroudsburg, PA, USA}, \bibinfo{pages}{18--27}.
\newblock


\bibitem[\protect\citeauthoryear{Arif and Stuerzlinger}{Arif and
  Stuerzlinger}{2010}]%
        {SAB+10}
\bibfield{author}{\bibinfo{person}{Ahmed~Sabbir Arif} {and}
  \bibinfo{person}{Wolfgang Stuerzlinger}.} \bibinfo{year}{2010}\natexlab{}.
\newblock \showarticletitle{Predicting the Cost of Error Correction in
  Character-Based Text Entry Technologies}. In
  \bibinfo{booktitle}{\emph{Proceedings of the SIGCHI Conference on Human
  Factors in Computing Systems}} (Atlanta, Georgia, USA)
  \emph{(\bibinfo{series}{CHI '10})}. \bibinfo{publisher}{Association for
  Computing Machinery}, \bibinfo{address}{New York, NY, USA},
  \bibinfo{pages}{5–14}.
\newblock
\showISBNx{9781605589299}
\urldef\tempurl%
\url{https://doi.org/10.1145/1753326.1753329}
\showDOI{\tempurl}


\bibitem[\protect\citeauthoryear{Baig and Kavakli}{Baig and Kavakli}{2018}]%
        {BAI+18}
\bibfield{author}{\bibinfo{person}{Muhammad~Zeeshan Baig} {and}
  \bibinfo{person}{Manolya Kavakli}.} \bibinfo{year}{2018}\natexlab{}.
\newblock \showarticletitle{Qualitative analysis of a multimodal interface
  system using speech/gesture}. In \bibinfo{booktitle}{\emph{2018 13th IEEE
  Conference on Industrial Electronics and Applications (ICIEA)}}. IEEE,
  \bibinfo{publisher}{IEEE}, \bibinfo{address}{Wuhan, China},
  \bibinfo{pages}{2811--2816}.
\newblock


\bibitem[\protect\citeauthoryear{Bolt}{Bolt}{1980}]%
        {BOL80}
\bibfield{author}{\bibinfo{person}{Richard~A. Bolt}.}
  \bibinfo{year}{1980}\natexlab{}.
\newblock \showarticletitle{\&Ldquo;Put-that-there\&Rdquo;: Voice and Gesture
  at the Graphics Interface}.
\newblock \bibinfo{journal}{\emph{SIGGRAPH Comput. Graph.}}
  \bibinfo{volume}{14}, \bibinfo{number}{3} (\bibinfo{date}{July}
  \bibinfo{year}{1980}), \bibinfo{pages}{262--270}.
\newblock
\showISSN{0097-8930}
\urldef\tempurl%
\url{https://doi.org/10.1145/965105.807503}
\showDOI{\tempurl}


\bibitem[\protect\citeauthoryear{Bourguet}{Bourguet}{2006}]%
        {BOU06}
\bibfield{author}{\bibinfo{person}{Marie-Luce Bourguet}.}
  \bibinfo{year}{2006}\natexlab{}.
\newblock \showarticletitle{Towards a taxonomy of error-handling strategies in
  recognition-based multi-modal human--computer interfaces}.
\newblock \bibinfo{journal}{\emph{Signal Processing}} \bibinfo{volume}{86},
  \bibinfo{number}{12} (\bibinfo{year}{2006}), \bibinfo{pages}{3625--3643}.
\newblock


\bibitem[\protect\citeauthoryear{Bourguet and Ando}{Bourguet and Ando}{1998}]%
        {BOU+98}
\bibfield{author}{\bibinfo{person}{Marie-Luce Bourguet} {and}
  \bibinfo{person}{Akio Ando}.} \bibinfo{year}{1998}\natexlab{}.
\newblock \showarticletitle{Synchronization of Speech and Hand Gestures during
  Multimodal Human-Computer Interaction}. In \bibinfo{booktitle}{\emph{CHI 98
  Conference Summary on Human Factors in Computing Systems}} (Los Angeles,
  California, USA) \emph{(\bibinfo{series}{CHI ’98})}.
  \bibinfo{publisher}{Association for Computing Machinery},
  \bibinfo{address}{New York, NY, USA}, \bibinfo{pages}{241–242}.
\newblock
\showISBNx{1581130287}
\urldef\tempurl%
\url{https://doi.org/10.1145/286498.286726}
\showDOI{\tempurl}


\bibitem[\protect\citeauthoryear{Bowman, Kruijff, LaViola, and Poupyrev}{Bowman
  et~al\mbox{.}}{2004}]%
        {3duibook}
\bibfield{author}{\bibinfo{person}{Doug~A. Bowman}, \bibinfo{person}{Ernst
  Kruijff}, \bibinfo{person}{Joseph~J. LaViola}, {and} \bibinfo{person}{Ivan
  Poupyrev}.} \bibinfo{year}{2004}\natexlab{}.
\newblock \bibinfo{booktitle}{\emph{3D User Interfaces: Theory and Practice}}.
\newblock \bibinfo{publisher}{Addison Wesley Longman Publishing Co., Inc.},
  \bibinfo{address}{USA}.
\newblock
\showISBNx{0201758679}


\bibitem[\protect\citeauthoryear{Brustein}{Brustein}{2018}]%
        {DOD}
\bibfield{author}{\bibinfo{person}{Joshua Brustein}.}
  \bibinfo{year}{2018}\natexlab{}.
\newblock \bibinfo{title}{Microsoft Wins \$480 Million Army Battlefield
  Contract}.
\newblock
\newblock
\urldef\tempurl%
\url{https://www.bloomberg.com/news/articles/2018-11-28/microsoft-wins-480-million-army-battlefield-contract}
\showURL{%
\tempurl}


\bibitem[\protect\citeauthoryear{Buchanan, Floyd, Holderness, and
  LaViola}{Buchanan et~al\mbox{.}}{2013}]%
        {BUC+13}
\bibfield{author}{\bibinfo{person}{Sarah Buchanan}, \bibinfo{person}{Bourke
  Floyd}, \bibinfo{person}{Will Holderness}, {and} \bibinfo{person}{Joseph~J.
  LaViola}.} \bibinfo{year}{2013}\natexlab{}.
\newblock \showarticletitle{Towards User-Defined Multi-Touch Gestures for 3D
  Objects}. In \bibinfo{booktitle}{\emph{Proceedings of the 2013 ACM
  International Conference on Interactive Tabletops and Surfaces}} (St.
  Andrews, Scotland, United Kingdom) \emph{(\bibinfo{series}{ITS ’13})}.
  \bibinfo{publisher}{Association for Computing Machinery},
  \bibinfo{address}{New York, NY, USA}, \bibinfo{pages}{231–240}.
\newblock
\showISBNx{9781450322713}
\urldef\tempurl%
\url{https://doi.org/10.1145/2512349.2512825}
\showDOI{\tempurl}


\bibitem[\protect\citeauthoryear{Carbini, Delphin-Poulat, Perron, and
  Viallet}{Carbini et~al\mbox{.}}{2006}]%
        {CAR+06}
\bibfield{author}{\bibinfo{person}{S{\'e}bastien Carbini},
  \bibinfo{person}{Lionel Delphin-Poulat}, \bibinfo{person}{L Perron}, {and}
  \bibinfo{person}{Jean-Emmanuel Viallet}.} \bibinfo{year}{2006}\natexlab{}.
\newblock \showarticletitle{From a wizard of Oz experiment to a real time
  speech and gesture multimodal interface}.
\newblock \bibinfo{journal}{\emph{Signal Processing}} \bibinfo{volume}{86},
  \bibinfo{number}{12} (\bibinfo{year}{2006}), \bibinfo{pages}{3559--3577}.
\newblock


\bibitem[\protect\citeauthoryear{Chai and Qu}{Chai and Qu}{2005}]%
        {EIS+04}
\bibfield{author}{\bibinfo{person}{Joyce~Y. Chai} {and}
  \bibinfo{person}{Shaolin Qu}.} \bibinfo{year}{2005}\natexlab{}.
\newblock \showarticletitle{A Salience Driven Approach to Robust Input
  Interpretation in Multimodal Conversational Systems}. In
  \bibinfo{booktitle}{\emph{Proceedings of the Conference on Human Language
  Technology and Empirical Methods in Natural Language Processing}} (Vancouver,
  British Columbia, Canada) \emph{(\bibinfo{series}{HLT ’05})}.
  \bibinfo{publisher}{Association for Computational Linguistics},
  \bibinfo{address}{USA}, \bibinfo{pages}{217–224}.
\newblock
\urldef\tempurl%
\url{https://doi.org/10.3115/1220575.1220603}
\showDOI{\tempurl}


\bibitem[\protect\citeauthoryear{Chan, Seyed, Stuerzlinger, Yang, and
  Maurer}{Chan et~al\mbox{.}}{2016}]%
        {CHA+16}
\bibfield{author}{\bibinfo{person}{Edwin Chan}, \bibinfo{person}{Teddy Seyed},
  \bibinfo{person}{Wolfgang Stuerzlinger}, \bibinfo{person}{Xing-Dong Yang},
  {and} \bibinfo{person}{Frank Maurer}.} \bibinfo{year}{2016}\natexlab{}.
\newblock \showarticletitle{User Elicitation on Single-Hand Microgestures}. In
  \bibinfo{booktitle}{\emph{Proceedings of the 2016 CHI Conference on Human
  Factors in Computing Systems}} (San Jose, California, USA)
  \emph{(\bibinfo{series}{CHI ’16})}. \bibinfo{publisher}{Association for
  Computing Machinery}, \bibinfo{address}{New York, NY, USA},
  \bibinfo{pages}{3403–3414}.
\newblock
\showISBNx{9781450333627}
\urldef\tempurl%
\url{https://doi.org/10.1145/2858036.2858589}
\showDOI{\tempurl}


\bibitem[\protect\citeauthoryear{Coh{\'e} and Hachet}{Coh{\'e} and
  Hachet}{2012}]%
        {COH+12}
\bibfield{author}{\bibinfo{person}{Aur{\'e}lie Coh{\'e}} {and}
  \bibinfo{person}{Martin Hachet}.} \bibinfo{year}{2012}\natexlab{}.
\newblock \showarticletitle{Understanding User Gestures for Manipulating 3D
  Objects from Touchscreen Inputs}. In \bibinfo{booktitle}{\emph{Proceedings of
  Graphics Interface 2012}} (Toronto, Ontario, Canada)
  \emph{(\bibinfo{series}{GI '12})}. \bibinfo{publisher}{Canadian Information
  Processing Society}, \bibinfo{address}{Toronto, Ont., Canada, Canada},
  \bibinfo{pages}{157--164}.
\newblock
\showISBNx{978-1-4503-1420-6}
\urldef\tempurl%
\url{http://dl.acm.org/citation.cfm?id=2305276.2305303}
\showURL{%
\tempurl}


\bibitem[\protect\citeauthoryear{Corradini and Cohen}{Corradini and
  Cohen}{2005}]%
        {COR+05}
\bibfield{author}{\bibinfo{person}{Andrea Corradini} {and}
  \bibinfo{person}{Philip~R Cohen}.} \bibinfo{year}{2005}\natexlab{}.
\newblock \bibinfo{title}{On the Relationships Among Speech, Gestures, and
  Object Manipulation in Virtual Environments: Initial Evidence}.
\newblock , \bibinfo{numpages}{97--112}~pages.
\newblock


\bibitem[\protect\citeauthoryear{D{\"u}nser, Grasset, Seichter, and
  Billinghurst}{D{\"u}nser et~al\mbox{.}}{2007}]%
        {BIL07}
\bibfield{author}{\bibinfo{person}{Andreas D{\"u}nser},
  \bibinfo{person}{Rapha{\"e}l Grasset}, \bibinfo{person}{Hartmut Seichter},
  {and} \bibinfo{person}{Mark Billinghurst}.} \bibinfo{year}{2007}\natexlab{}.
\newblock \bibinfo{booktitle}{\emph{Applying HCI principles to AR systems
  design}}.
\newblock \bibinfo{publisher}{University of Canterbury. Human Interface
  Technology Laboratory.}, \bibinfo{address}{New Zealand}.
\newblock


\bibitem[\protect\citeauthoryear{Goldin-Meadow, Alibali, and
  Church}{Goldin-Meadow et~al\mbox{.}}{1993}]%
        {GOL+93}
\bibfield{author}{\bibinfo{person}{Susan Goldin-Meadow},
  \bibinfo{person}{Martha~Wagner Alibali}, {and}
  \bibinfo{person}{R~Breckinridge Church}.} \bibinfo{year}{1993}\natexlab{}.
\newblock \showarticletitle{Transitions in concept acquisition: using the hand
  to read the mind.}
\newblock \bibinfo{journal}{\emph{Psychological review}} \bibinfo{volume}{100},
  \bibinfo{number}{2} (\bibinfo{year}{1993}), \bibinfo{pages}{279}.
\newblock


\bibitem[\protect\citeauthoryear{Harada, Sato, Takagi, and Asakawa}{Harada
  et~al\mbox{.}}{2013}]%
        {SUS+13}
\bibfield{author}{\bibinfo{person}{Susumu Harada}, \bibinfo{person}{Daisuke
  Sato}, \bibinfo{person}{Hironobu Takagi}, {and} \bibinfo{person}{Chieko
  Asakawa}.} \bibinfo{year}{2013}\natexlab{}.
\newblock \showarticletitle{Characteristics of Elderly User Behavior on Mobile
  Multi-touch Devices}. In \bibinfo{booktitle}{\emph{Human-Computer Interaction
  -- INTERACT 2013}}, \bibfield{editor}{\bibinfo{person}{Paula Kotz{\'e}},
  \bibinfo{person}{Gary Marsden}, \bibinfo{person}{Gitte Lindgaard},
  \bibinfo{person}{Janet Wesson}, {and} \bibinfo{person}{Marco Winckler}}
  (Eds.). \bibinfo{publisher}{Springer Berlin Heidelberg},
  \bibinfo{address}{Berlin, Heidelberg}, \bibinfo{pages}{323--341}.
\newblock
\showISBNx{978-3-642-40498-6}


\bibitem[\protect\citeauthoryear{Hauptmann}{Hauptmann}{1989}]%
        {HAU89}
\bibfield{author}{\bibinfo{person}{A~G Hauptmann}.}
  \bibinfo{year}{1989}\natexlab{}.
\newblock \showarticletitle{Speech and gestures for graphic image
  manipulation}.
\newblock \bibinfo{journal}{\emph{ACM SIGCHI Bulletin}} \bibinfo{volume}{20},
  \bibinfo{number}{SI} (\bibinfo{year}{1989}), \bibinfo{pages}{241--245}.
\newblock


\bibitem[\protect\citeauthoryear{Hauptmann and McAvinney}{Hauptmann and
  McAvinney}{1993}]%
        {HAU+93}
\bibfield{author}{\bibinfo{person}{Alexander~G Hauptmann} {and}
  \bibinfo{person}{Paul McAvinney}.} \bibinfo{year}{1993}\natexlab{}.
\newblock \showarticletitle{Gestures with speech for graphic manipulation}.
\newblock \bibinfo{journal}{\emph{International Journal of Man-Machine
  Studies}} \bibinfo{volume}{38}, \bibinfo{number}{2} (\bibinfo{year}{1993}),
  \bibinfo{pages}{231--249}.
\newblock


\bibitem[\protect\citeauthoryear{Irawati, Green, Billinghurst, Duenser, and
  Ko}{Irawati et~al\mbox{.}}{2006}]%
        {IRA+06}
\bibfield{author}{\bibinfo{person}{Sylvia Irawati}, \bibinfo{person}{Scott
  Green}, \bibinfo{person}{Mark Billinghurst}, \bibinfo{person}{Andreas
  Duenser}, {and} \bibinfo{person}{Heedong Ko}.}
  \bibinfo{year}{2006}\natexlab{}.
\newblock \showarticletitle{An Evaluation of an Augmented Reality Multimodal
  Interface Using Speech and Paddle Gestures}. In
  \bibinfo{booktitle}{\emph{Proceedings of the 16th International Conference on
  Advances in Artificial Reality and Tele-Existence}} (Hangzhou, China)
  \emph{(\bibinfo{series}{ICAT’06})}. \bibinfo{publisher}{Springer-Verlag},
  \bibinfo{address}{Berlin, Heidelberg}, \bibinfo{pages}{272–283}.
\newblock
\showISBNx{3540497765}
\urldef\tempurl%
\url{https://doi.org/10.1007/11941354_28}
\showDOI{\tempurl}


\bibitem[\protect\citeauthoryear{Jacob, Girouard, Hirshfield, Horn, Shaer,
  Solovey, and Zigelbaum}{Jacob et~al\mbox{.}}{2008}]%
        {JAC+08}
\bibfield{author}{\bibinfo{person}{Robert~J.K. Jacob}, \bibinfo{person}{Audrey
  Girouard}, \bibinfo{person}{Leanne~M. Hirshfield},
  \bibinfo{person}{Michael~S. Horn}, \bibinfo{person}{Orit Shaer},
  \bibinfo{person}{Erin~Treacy Solovey}, {and} \bibinfo{person}{Jamie
  Zigelbaum}.} \bibinfo{year}{2008}\natexlab{}.
\newblock \showarticletitle{Reality-Based Interaction: A Framework for
  Post-WIMP Interfaces}. In \bibinfo{booktitle}{\emph{Proceedings of the SIGCHI
  Conference on Human Factors in Computing Systems}} (Florence, Italy)
  \emph{(\bibinfo{series}{CHI '08})}. \bibinfo{publisher}{Association for
  Computing Machinery}, \bibinfo{address}{New York, NY, USA},
  \bibinfo{pages}{201–210}.
\newblock
\showISBNx{9781605580111}
\urldef\tempurl%
\url{https://doi.org/10.1145/1357054.1357089}
\showDOI{\tempurl}


\bibitem[\protect\citeauthoryear{Johnston, Cohen, McGee, Oviatt, Pittman, and
  Smith}{Johnston et~al\mbox{.}}{1997}]%
        {JOH+97}
\bibfield{author}{\bibinfo{person}{Michael Johnston},
  \bibinfo{person}{Philip~R. Cohen}, \bibinfo{person}{David McGee},
  \bibinfo{person}{Sharon~L. Oviatt}, \bibinfo{person}{James~A. Pittman}, {and}
  \bibinfo{person}{Ira Smith}.} \bibinfo{year}{1997}\natexlab{}.
\newblock \showarticletitle{Unification-Based Multimodal Integration}. In
  \bibinfo{booktitle}{\emph{Proceedings of the 35th Annual Meeting of the
  Association for Computational Linguistics and Eighth Conference of the
  European Chapter of the Association for Computational Linguistics}} (Madrid,
  Spain) \emph{(\bibinfo{series}{ACL ’98/EACL ’98})}.
  \bibinfo{publisher}{Association for Computational Linguistics},
  \bibinfo{address}{USA}, \bibinfo{pages}{281–288}.
\newblock
\urldef\tempurl%
\url{https://doi.org/10.3115/976909.979653}
\showDOI{\tempurl}


\bibitem[\protect\citeauthoryear{Kaiser, Olwal, McGee, Benko, Corradini, Li,
  Cohen, and Feiner}{Kaiser et~al\mbox{.}}{2003}]%
        {KAI+03}
\bibfield{author}{\bibinfo{person}{Ed Kaiser}, \bibinfo{person}{Alex Olwal},
  \bibinfo{person}{David McGee}, \bibinfo{person}{Hrvoje Benko},
  \bibinfo{person}{Andrea Corradini}, \bibinfo{person}{Xiaoguang Li},
  \bibinfo{person}{Phil Cohen}, {and} \bibinfo{person}{Steven Feiner}.}
  \bibinfo{year}{2003}\natexlab{}.
\newblock \showarticletitle{Mutual Disambiguation of 3D Multimodal Interaction
  in Augmented and Virtual Reality}. In \bibinfo{booktitle}{\emph{Proceedings
  of the 5th International Conference on Multimodal Interfaces}} (Vancouver,
  British Columbia, Canada) \emph{(\bibinfo{series}{ICMI ’03})}.
  \bibinfo{publisher}{Association for Computing Machinery},
  \bibinfo{address}{New York, NY, USA}, \bibinfo{pages}{12–19}.
\newblock
\showISBNx{1581136218}
\urldef\tempurl%
\url{https://doi.org/10.1145/958432.958438}
\showDOI{\tempurl}


\bibitem[\protect\citeauthoryear{Karpov and Yusupov}{Karpov and
  Yusupov}{2018}]%
        {KAR+18}
\bibfield{author}{\bibinfo{person}{A~A Karpov} {and} \bibinfo{person}{R~M
  Yusupov}.} \bibinfo{year}{2018}\natexlab{}.
\newblock \showarticletitle{Multimodal Interfaces of {Human--Computer}
  Interaction}.
\newblock \bibinfo{journal}{\emph{Her. Russ. Acad. Sci.}} \bibinfo{volume}{88},
  \bibinfo{number}{1} (\bibinfo{date}{Jan.} \bibinfo{year}{2018}),
  \bibinfo{pages}{67--74}.
\newblock


\bibitem[\protect\citeauthoryear{Kelly, Ozy{\"u}rek, and Maris}{Kelly
  et~al\mbox{.}}{2010}]%
        {KEL+10}
\bibfield{author}{\bibinfo{person}{Spencer~D Kelly}, \bibinfo{person}{Asli
  Ozy{\"u}rek}, {and} \bibinfo{person}{Eric Maris}.}
  \bibinfo{year}{2010}\natexlab{}.
\newblock \showarticletitle{Two sides of the same coin: speech and gesture
  mutually interact to enhance comprehension}.
\newblock \bibinfo{journal}{\emph{Psychol. Sci.}} \bibinfo{volume}{21},
  \bibinfo{number}{2} (\bibinfo{date}{Feb.} \bibinfo{year}{2010}),
  \bibinfo{pages}{260--267}.
\newblock


\bibitem[\protect\citeauthoryear{Khan and Tun{\c{c}}er}{Khan and
  Tun{\c{c}}er}{2019}]%
        {KHA+19}
\bibfield{author}{\bibinfo{person}{Sumbul Khan} {and} \bibinfo{person}{Bige
  Tun{\c{c}}er}.} \bibinfo{year}{2019}\natexlab{}.
\newblock \showarticletitle{Gesture and speech elicitation for 3D CAD modeling
  in conceptual design}.
\newblock \bibinfo{journal}{\emph{Automation in Construction}}
  \bibinfo{volume}{106} (\bibinfo{year}{2019}), \bibinfo{pages}{102847}.
\newblock


\bibitem[\protect\citeauthoryear{Kin, Agrawala, and DeRose}{Kin
  et~al\mbox{.}}{2009}]%
        {KIN+09}
\bibfield{author}{\bibinfo{person}{Kenrick Kin}, \bibinfo{person}{Maneesh
  Agrawala}, {and} \bibinfo{person}{Tony DeRose}.}
  \bibinfo{year}{2009}\natexlab{}.
\newblock \showarticletitle{Determining the Benefits of Direct-Touch, Bimanual,
  and Multifinger Input on a Multitouch Workstation}. In
  \bibinfo{booktitle}{\emph{Proceedings of Graphics Interface 2009}} (Kelowna,
  British Columbia, Canada) \emph{(\bibinfo{series}{GI '09})}.
  \bibinfo{publisher}{Canadian Information Processing Society},
  \bibinfo{address}{CAN}, \bibinfo{pages}{119–124}.
\newblock
\showISBNx{9781568814704}


\bibitem[\protect\citeauthoryear{Koons, Sparrell, and Thorisson}{Koons
  et~al\mbox{.}}{1998}]%
        {KOO+93}
\bibfield{author}{\bibinfo{person}{David~B. Koons}, \bibinfo{person}{Carlton~J.
  Sparrell}, {and} \bibinfo{person}{Kristinn~Rr. Thorisson}.}
  \bibinfo{year}{1998}\natexlab{}.
\newblock \bibinfo{booktitle}{\emph{Integrating Simultaneous Input from Speech,
  Gaze, and Hand Gestures}}.
\newblock \bibinfo{publisher}{Morgan Kaufmann Publishers Inc.},
  \bibinfo{address}{San Francisco, CA, USA}, \bibinfo{pages}{53–64}.
\newblock
\showISBNx{1558604448}


\bibitem[\protect\citeauthoryear{Lee and Billinghurst}{Lee and
  Billinghurst}{2008}]%
        {LEE+08}
\bibfield{author}{\bibinfo{person}{Minkyung Lee} {and} \bibinfo{person}{Mark
  Billinghurst}.} \bibinfo{year}{2008}\natexlab{}.
\newblock \showarticletitle{A Wizard of Oz Study for an AR Multimodal
  Interface}. In \bibinfo{booktitle}{\emph{Proceedings of the 10th
  International Conference on Multimodal Interfaces}} (Chania, Crete, Greece)
  \emph{(\bibinfo{series}{ICMI ’08})}. \bibinfo{publisher}{Association for
  Computing Machinery}, \bibinfo{address}{New York, NY, USA},
  \bibinfo{pages}{249–256}.
\newblock
\showISBNx{9781605581989}
\urldef\tempurl%
\url{https://doi.org/10.1145/1452392.1452444}
\showDOI{\tempurl}


\bibitem[\protect\citeauthoryear{Lee, Billinghurst, Baek, Green, and Woo}{Lee
  et~al\mbox{.}}{2013}]%
        {LEE+13}
\bibfield{author}{\bibinfo{person}{Minkyung Lee}, \bibinfo{person}{Mark
  Billinghurst}, \bibinfo{person}{Woonhyuk Baek}, \bibinfo{person}{Richard
  Green}, {and} \bibinfo{person}{Woontack Woo}.}
  \bibinfo{year}{2013}\natexlab{}.
\newblock \showarticletitle{A usability study of multimodal input in an
  augmented reality environment}.
\newblock \bibinfo{journal}{\emph{Virtual Real.}} \bibinfo{volume}{17},
  \bibinfo{number}{4} (\bibinfo{date}{Nov.} \bibinfo{year}{2013}),
  \bibinfo{pages}{293--305}.
\newblock


\bibitem[\protect\citeauthoryear{Loehr}{Loehr}{2012}]%
        {LOE12}
\bibfield{author}{\bibinfo{person}{Daniel~P Loehr}.}
  \bibinfo{year}{2012}\natexlab{}.
\newblock \showarticletitle{Temporal, structural, and pragmatic synchrony
  between intonation and gesture}.
\newblock \bibinfo{journal}{\emph{Laboratory Phonology}} \bibinfo{volume}{3},
  \bibinfo{number}{1} (\bibinfo{year}{2012}), \bibinfo{pages}{71--89}.
\newblock


\bibitem[\protect\citeauthoryear{Mcneill}{Mcneill}{2005}]%
        {MCN05}
\bibfield{author}{\bibinfo{person}{David Mcneill}.}
  \bibinfo{year}{2005}\natexlab{}.
\newblock \bibinfo{booktitle}{\emph{Gesture and Thought}}.
\newblock \bibinfo{publisher}{the University of Chicago Press},
  \bibinfo{address}{USA}.
\newblock
\urldef\tempurl%
\url{https://doi.org/10.7208/chicago/9780226514642.001.0001}
\showDOI{\tempurl}


\bibitem[\protect\citeauthoryear{Micire, Desai, Courtemanche, Tsui, and
  Yanco}{Micire et~al\mbox{.}}{2009}]%
        {MIC+09}
\bibfield{author}{\bibinfo{person}{Mark Micire}, \bibinfo{person}{Munjal
  Desai}, \bibinfo{person}{Amanda Courtemanche}, \bibinfo{person}{Katherine~M.
  Tsui}, {and} \bibinfo{person}{Holly~A. Yanco}.}
  \bibinfo{year}{2009}\natexlab{}.
\newblock \showarticletitle{Analysis of Natural Gestures for Controlling Robot
  Teams on Multi-touch Tabletop Surfaces}. In
  \bibinfo{booktitle}{\emph{Proceedings of the ACM International Conference on
  Interactive Tabletops and Surfaces}} (Banff, Alberta, Canada)
  \emph{(\bibinfo{series}{ITS '09})}. \bibinfo{publisher}{ACM},
  \bibinfo{address}{New York, NY, USA}, \bibinfo{pages}{41--48}.
\newblock
\showISBNx{978-1-60558-733-2}
\urldef\tempurl%
\url{https://doi.org/10.1145/1731903.1731912}
\showDOI{\tempurl}


\bibitem[\protect\citeauthoryear{Mignot, Valot, and Carbonell}{Mignot
  et~al\mbox{.}}{1993}]%
        {MIG+93}
\bibfield{author}{\bibinfo{person}{Christophe Mignot}, \bibinfo{person}{Claude
  Valot}, {and} \bibinfo{person}{No\"{e}lle Carbonell}.}
  \bibinfo{year}{1993}\natexlab{}.
\newblock \showarticletitle{An Experimental Study of Future “Natural”
  Multimodal Human-Computer Interaction}. In \bibinfo{booktitle}{\emph{INTERACT
  ’93 and CHI ’93 Conference Companion on Human Factors in Computing
  Systems}} (Amsterdam, The Netherlands) \emph{(\bibinfo{series}{CHI ’93})}.
  \bibinfo{publisher}{Association for Computing Machinery},
  \bibinfo{address}{New York, NY, USA}, \bibinfo{pages}{67–68}.
\newblock
\showISBNx{0897915747}
\urldef\tempurl%
\url{https://doi.org/10.1145/259964.260075}
\showDOI{\tempurl}


\bibitem[\protect\citeauthoryear{Mol and Kita}{Mol and Kita}{2012}]%
        {MOL+12}
\bibfield{author}{\bibinfo{person}{Lisette Mol} {and} \bibinfo{person}{Sotaro
  Kita}.} \bibinfo{year}{2012}\natexlab{}.
\newblock \showarticletitle{Gesture structure affects syntactic structure in
  speech}. In \bibinfo{booktitle}{\emph{Proceedings of the Annual Meeting of
  the Cognitive Science Society}}, Vol.~\bibinfo{volume}{34}.
  \bibinfo{publisher}{CogSci}, \bibinfo{address}{USA}, \bibinfo{pages}{761 --
  766}.
\newblock


\bibitem[\protect\citeauthoryear{Morris}{Morris}{2012}]%
        {MOR12}
\bibfield{author}{\bibinfo{person}{Meredith~Ringel Morris}.}
  \bibinfo{year}{2012}\natexlab{}.
\newblock \showarticletitle{Web on the Wall: Insights from a Multimodal
  Interaction Elicitation Study}. In \bibinfo{booktitle}{\emph{Proceedings of
  the 2012 ACM International Conference on Interactive Tabletops and Surfaces}}
  (Cambridge, Massachusetts, USA) \emph{(\bibinfo{series}{ITS '12})}.
  \bibinfo{publisher}{ACM}, \bibinfo{address}{New York, NY, USA},
  \bibinfo{pages}{95--104}.
\newblock
\showISBNx{978-1-4503-1209-7}
\urldef\tempurl%
\url{https://doi.org/10.1145/2396636.2396651}
\showDOI{\tempurl}


\bibitem[\protect\citeauthoryear{Morris, Wobbrock, and Wilson}{Morris
  et~al\mbox{.}}{2010}]%
        {MOR+10}
\bibfield{author}{\bibinfo{person}{Meredith~Ringel Morris},
  \bibinfo{person}{Jacob~O. Wobbrock}, {and} \bibinfo{person}{Andrew~D.
  Wilson}.} \bibinfo{year}{2010}\natexlab{}.
\newblock \showarticletitle{Understanding Users' Preferences for Surface
  Gestures}. In \bibinfo{booktitle}{\emph{Proceedings of Graphics Interface
  2010}} (Ottawa, Ontario, Canada) \emph{(\bibinfo{series}{GI '10})}.
  \bibinfo{publisher}{Canadian Information Processing Society},
  \bibinfo{address}{CAN}, \bibinfo{pages}{261–268}.
\newblock
\showISBNx{9781568817125}


\bibitem[\protect\citeauthoryear{Moscovich and Hughes}{Moscovich and
  Hughes}{2008}]%
        {MOS+08}
\bibfield{author}{\bibinfo{person}{Tomer Moscovich} {and}
  \bibinfo{person}{John~F. Hughes}.} \bibinfo{year}{2008}\natexlab{}.
\newblock \showarticletitle{Indirect Mappings of Multi-Touch Input Using One
  and Two Hands}. In \bibinfo{booktitle}{\emph{Proceedings of the SIGCHI
  Conference on Human Factors in Computing Systems}} (Florence, Italy)
  \emph{(\bibinfo{series}{CHI '08})}. \bibinfo{publisher}{Association for
  Computing Machinery}, \bibinfo{address}{New York, NY, USA},
  \bibinfo{pages}{1275–1284}.
\newblock
\showISBNx{9781605580111}
\urldef\tempurl%
\url{https://doi.org/10.1145/1357054.1357254}
\showDOI{\tempurl}


\bibitem[\protect\citeauthoryear{Nacenta, Kamber, Qiang, and
  Kristensson}{Nacenta et~al\mbox{.}}{2013}]%
        {NAC+13}
\bibfield{author}{\bibinfo{person}{Miguel~A Nacenta}, \bibinfo{person}{Yemliha
  Kamber}, \bibinfo{person}{Yizhou Qiang}, {and} \bibinfo{person}{Per~Ola
  Kristensson}.} \bibinfo{year}{2013}\natexlab{}.
\newblock \bibinfo{title}{Memorability of pre-designed and user-defined gesture
  sets}.
\newblock
\newblock


\bibitem[\protect\citeauthoryear{Nielsen, St{\"o}rring, Moeslund, and
  Granum}{Nielsen et~al\mbox{.}}{2004}]%
        {NIE+04}
\bibfield{author}{\bibinfo{person}{Michael Nielsen}, \bibinfo{person}{Moritz
  St{\"o}rring}, \bibinfo{person}{Thomas~B. Moeslund}, {and}
  \bibinfo{person}{Erik Granum}.} \bibinfo{year}{2004}\natexlab{}.
\newblock \showarticletitle{A Procedure for Developing Intuitive and Ergonomic
  Gesture Interfaces for HCI}. In \bibinfo{booktitle}{\emph{Gesture-Based
  Communication in Human-Computer Interaction}},
  \bibfield{editor}{\bibinfo{person}{Antonio Camurri} {and}
  \bibinfo{person}{Gualtiero Volpe}} (Eds.). \bibinfo{publisher}{Springer
  Berlin Heidelberg}, \bibinfo{address}{Berlin, Heidelberg},
  \bibinfo{pages}{409--420}.
\newblock
\showISBNx{978-3-540-24598-8}


\bibitem[\protect\citeauthoryear{{Ortega}, {Galvan}, {Tarre}, {Barreto},
  {Rishe}, {Bernal}, {Balcazar}, and {Thomas}}{{Ortega} et~al\mbox{.}}{2017}]%
        {ORT+17}
\bibfield{author}{\bibinfo{person}{F.~R. {Ortega}}, \bibinfo{person}{A.
  {Galvan}}, \bibinfo{person}{K. {Tarre}}, \bibinfo{person}{A. {Barreto}},
  \bibinfo{person}{N. {Rishe}}, \bibinfo{person}{J. {Bernal}},
  \bibinfo{person}{R. {Balcazar}}, {and} \bibinfo{person}{J. {Thomas}}.}
  \bibinfo{year}{2017}\natexlab{}.
\newblock \showarticletitle{Gesture elicitation for 3D travel via multi-touch
  and mid-Air systems for procedurally generated pseudo-universe}. In
  \bibinfo{booktitle}{\emph{2017 IEEE Symposium on 3D User Interfaces (3DUI)}}.
  \bibinfo{publisher}{IEEE}, \bibinfo{address}{Los Angeles, CA, USA},
  \bibinfo{pages}{144--153}.
\newblock


\bibitem[\protect\citeauthoryear{{Ortega}, {Tarre}, {Kress}, {Williams},
  {Barreto}, and {Rishe}}{{Ortega} et~al\mbox{.}}{2019}]%
        {ORT+19}
\bibfield{author}{\bibinfo{person}{F.~R. {Ortega}}, \bibinfo{person}{K.
  {Tarre}}, \bibinfo{person}{M. {Kress}}, \bibinfo{person}{A.~S. {Williams}},
  \bibinfo{person}{A.~B. {Barreto}}, {and} \bibinfo{person}{N.~D. {Rishe}}.}
  \bibinfo{year}{2019}\natexlab{}.
\newblock \showarticletitle{Selection and Manipulation Whole-Body Gesture
  Elicitation Study In Virtual Reality}. In \bibinfo{booktitle}{\emph{2019 IEEE
  Conference on Virtual Reality and 3D User Interfaces (VR)}}.
  \bibinfo{publisher}{IEEE}, \bibinfo{address}{Osaka, Japan, Japan},
  \bibinfo{pages}{1723--1728}.
\newblock


\bibitem[\protect\citeauthoryear{Oviatt}{Oviatt}{1999}]%
        {OVI99}
\bibfield{author}{\bibinfo{person}{Sharon Oviatt}.}
  \bibinfo{year}{1999}\natexlab{}.
\newblock \showarticletitle{Ten Myths of Multimodal Interaction}.
\newblock \bibinfo{journal}{\emph{Commun. ACM}} \bibinfo{volume}{42},
  \bibinfo{number}{11} (\bibinfo{date}{Nov.} \bibinfo{year}{1999}),
  \bibinfo{pages}{74–81}.
\newblock
\showISSN{0001-0782}
\urldef\tempurl%
\url{https://doi.org/10.1145/319382.319398}
\showDOI{\tempurl}


\bibitem[\protect\citeauthoryear{Oviatt}{Oviatt}{2000}]%
        {OVI00}
\bibfield{author}{\bibinfo{person}{Sharon Oviatt}.}
  \bibinfo{year}{2000}\natexlab{}.
\newblock \showarticletitle{Taming recognition errors with a multimodal
  interface}.
\newblock \bibinfo{journal}{\emph{Commun. ACM}} \bibinfo{volume}{43},
  \bibinfo{number}{9} (\bibinfo{year}{2000}), \bibinfo{pages}{45--51}.
\newblock


\bibitem[\protect\citeauthoryear{Oviatt, Coulston, and Lunsford}{Oviatt
  et~al\mbox{.}}{2004}]%
        {OVI+04}
\bibfield{author}{\bibinfo{person}{Sharon Oviatt}, \bibinfo{person}{Rachel
  Coulston}, {and} \bibinfo{person}{Rebecca Lunsford}.}
  \bibinfo{year}{2004}\natexlab{}.
\newblock \showarticletitle{When Do We Interact Multimodally? Cognitive Load
  and Multimodal Communication Patterns}. In
  \bibinfo{booktitle}{\emph{Proceedings of the 6th International Conference on
  Multimodal Interfaces}} (State College, PA, USA) \emph{(\bibinfo{series}{ICMI
  '04})}. \bibinfo{publisher}{Association for Computing Machinery},
  \bibinfo{address}{New York, NY, USA}, \bibinfo{pages}{129–136}.
\newblock
\showISBNx{1581139950}
\urldef\tempurl%
\url{https://doi.org/10.1145/1027933.1027957}
\showDOI{\tempurl}


\bibitem[\protect\citeauthoryear{Oviatt, DeAngeli, and Kuhn}{Oviatt
  et~al\mbox{.}}{1997}]%
        {OVI+97}
\bibfield{author}{\bibinfo{person}{Sharon Oviatt}, \bibinfo{person}{Antonella
  DeAngeli}, {and} \bibinfo{person}{Karen Kuhn}.}
  \bibinfo{year}{1997}\natexlab{}.
\newblock \showarticletitle{Integration and Synchronization of Input Modes
  during Multimodal Human-Computer Interaction}. In
  \bibinfo{booktitle}{\emph{Proceedings of the ACM SIGCHI Conference on Human
  Factors in Computing Systems}} (Atlanta, Georgia, USA)
  \emph{(\bibinfo{series}{CHI ’97})}. \bibinfo{publisher}{Association for
  Computing Machinery}, \bibinfo{address}{New York, NY, USA},
  \bibinfo{pages}{415–422}.
\newblock
\showISBNx{0897918029}
\urldef\tempurl%
\url{https://doi.org/10.1145/258549.258821}
\showDOI{\tempurl}


\bibitem[\protect\citeauthoryear{Petersson, Sinkvist, Wang, and
  Smedby}{Petersson et~al\mbox{.}}{2009}]%
        {PET+09}
\bibfield{author}{\bibinfo{person}{Helge Petersson}, \bibinfo{person}{David
  Sinkvist}, \bibinfo{person}{Chunliang Wang}, {and} \bibinfo{person}{{\"O}rjan
  Smedby}.} \bibinfo{year}{2009}\natexlab{}.
\newblock \showarticletitle{Web-based interactive 3D visualization as a tool
  for improved anatomy learning}.
\newblock \bibinfo{journal}{\emph{Anatomical sciences education}}
  \bibinfo{volume}{2}, \bibinfo{number}{2} (\bibinfo{year}{2009}),
  \bibinfo{pages}{61--68}.
\newblock


\bibitem[\protect\citeauthoryear{{Piumsomboon}, {Altimira}, {Kim}, {Clark},
  {Lee}, and {Billinghurst}}{{Piumsomboon} et~al\mbox{.}}{2014}]%
        {PIU+14}
\bibfield{author}{\bibinfo{person}{T. {Piumsomboon}}, \bibinfo{person}{D.
  {Altimira}}, \bibinfo{person}{H. {Kim}}, \bibinfo{person}{A. {Clark}},
  \bibinfo{person}{G. {Lee}}, {and} \bibinfo{person}{M. {Billinghurst}}.}
  \bibinfo{year}{2014}\natexlab{}.
\newblock \showarticletitle{Grasp-Shell vs gesture-speech: A comparison of
  direct and indirect natural interaction techniques in augmented reality}. In
  \bibinfo{booktitle}{\emph{2014 IEEE International Symposium on Mixed and
  Augmented Reality (ISMAR)}}. \bibinfo{publisher}{IEEE},
  \bibinfo{address}{Munich, Germany}, \bibinfo{pages}{73--82}.
\newblock


\bibitem[\protect\citeauthoryear{Piumsomboon, Clark, Billinghurst, and
  Cockburn}{Piumsomboon et~al\mbox{.}}{2013}]%
        {PIU+13}
\bibfield{author}{\bibinfo{person}{Thammathip Piumsomboon},
  \bibinfo{person}{Adrian Clark}, \bibinfo{person}{Mark Billinghurst}, {and}
  \bibinfo{person}{Andy Cockburn}.} \bibinfo{year}{2013}\natexlab{}.
\newblock \showarticletitle{User-Defined Gestures for Augmented Reality}. In
  \bibinfo{booktitle}{\emph{CHI ’13 Extended Abstracts on Human Factors in
  Computing Systems}} (Paris, France) \emph{(\bibinfo{series}{CHI EA ’13})}.
  \bibinfo{publisher}{Association for Computing Machinery},
  \bibinfo{address}{New York, NY, USA}, \bibinfo{pages}{955–960}.
\newblock
\showISBNx{9781450319522}
\urldef\tempurl%
\url{https://doi.org/10.1145/2468356.2468527}
\showDOI{\tempurl}


\bibitem[\protect\citeauthoryear{Plank, Jetter, R\"{a}dle, Klokmose, Luger, and
  Reiterer}{Plank et~al\mbox{.}}{2017}]%
        {PLA+17}
\bibfield{author}{\bibinfo{person}{Thomas Plank},
  \bibinfo{person}{Hans-Christian Jetter}, \bibinfo{person}{Roman R\"{a}dle},
  \bibinfo{person}{Clemens~N. Klokmose}, \bibinfo{person}{Thomas Luger}, {and}
  \bibinfo{person}{Harald Reiterer}.} \bibinfo{year}{2017}\natexlab{}.
\newblock \showarticletitle{Is Two Enough?: ! Studying Benefits, Barriers, and
  Biases of Multi-Tablet Use for Collaborative Visualization}. In
  \bibinfo{booktitle}{\emph{Proceedings of the 2017 CHI Conference on Human
  Factors in Computing Systems}} (Denver, Colorado, USA)
  \emph{(\bibinfo{series}{CHI '17})}. \bibinfo{publisher}{ACM},
  \bibinfo{address}{New York, NY, USA}, \bibinfo{pages}{4548--4560}.
\newblock
\showISBNx{978-1-4503-4655-9}
\urldef\tempurl%
\url{https://doi.org/10.1145/3025453.3025537}
\showDOI{\tempurl}


\bibitem[\protect\citeauthoryear{Robbe}{Robbe}{1998}]%
        {ROB98}
\bibfield{author}{\bibinfo{person}{Sandrine Robbe}.}
  \bibinfo{year}{1998}\natexlab{}.
\newblock \showarticletitle{An Empirical Study of Speech and Gesture
  Interaction: Toward the Definition of Ergonomic Design Guidelines}. In
  \bibinfo{booktitle}{\emph{CHI 98 Conference Summary on Human Factors in
  Computing Systems}} (Los Angeles, California, USA)
  \emph{(\bibinfo{series}{CHI ’98})}. \bibinfo{publisher}{Association for
  Computing Machinery}, \bibinfo{address}{New York, NY, USA},
  \bibinfo{pages}{349–350}.
\newblock
\showISBNx{1581130287}
\urldef\tempurl%
\url{https://doi.org/10.1145/286498.286815}
\showDOI{\tempurl}


\bibitem[\protect\citeauthoryear{Ruiz, Li, and Lank}{Ruiz
  et~al\mbox{.}}{2011}]%
        {RUI+11}
\bibfield{author}{\bibinfo{person}{Jaime Ruiz}, \bibinfo{person}{Yang Li},
  {and} \bibinfo{person}{Edward Lank}.} \bibinfo{year}{2011}\natexlab{}.
\newblock \bibinfo{title}{User-defined motion gestures for mobile interaction}.
\newblock
\newblock


\bibitem[\protect\citeauthoryear{Schegloff}{Schegloff}{1984}]%
        {SCH84}
\bibfield{author}{\bibinfo{person}{Emanuel~A Schegloff}.}
  \bibinfo{year}{1984}\natexlab{}.
\newblock \bibinfo{title}{On some gestures' relation to talk.(pp. 266-296) In
  J. Maxwell and J. Heritage (Eds.) Structures of social action}.
\newblock
\newblock


\bibitem[\protect\citeauthoryear{Tarre, Williams, Borges, Rishe, Barreto, and
  Ortega}{Tarre et~al\mbox{.}}{2018}]%
        {TAR+18}
\bibfield{author}{\bibinfo{person}{Katherine Tarre}, \bibinfo{person}{Adam~S.
  Williams}, \bibinfo{person}{Lukas Borges}, \bibinfo{person}{Naphtali~D.
  Rishe}, \bibinfo{person}{Armando~B. Barreto}, {and}
  \bibinfo{person}{Francisco~R. Ortega}.} \bibinfo{year}{2018}\natexlab{}.
\newblock \showarticletitle{Towards First Person Gamer Modeling and the Problem
  with Game Classification in User Studies}. In
  \bibinfo{booktitle}{\emph{Proceedings of the 24th ACM Symposium on Virtual
  Reality Software and Technology}} (Tokyo, Japan) \emph{(\bibinfo{series}{VRST
  '18})}. \bibinfo{publisher}{ACM}, \bibinfo{address}{New York, NY, USA},
  Article \bibinfo{articleno}{125}, \bibinfo{numpages}{2}~pages.
\newblock
\showISBNx{978-1-4503-6086-9}
\urldef\tempurl%
\url{https://doi.org/10.1145/3281505.3281590}
\showDOI{\tempurl}


\bibitem[\protect\citeauthoryear{Tsandilas}{Tsandilas}{2018}]%
        {TSA18}
\bibfield{author}{\bibinfo{person}{Theophanis Tsandilas}.}
  \bibinfo{year}{2018}\natexlab{}.
\newblock \showarticletitle{Fallacies of Agreement: A Critical Review of
  Consensus Assessment Methods for Gesture Elicitation}.
\newblock \bibinfo{journal}{\emph{ACM Trans. Comput. Hum. Interact.}}
  \bibinfo{volume}{25}, \bibinfo{number}{3} (\bibinfo{date}{June}
  \bibinfo{year}{2018}), \bibinfo{pages}{18}.
\newblock


\bibitem[\protect\citeauthoryear{Vatavu and Wobbrock}{Vatavu and
  Wobbrock}{2015}]%
        {VAT+15}
\bibfield{author}{\bibinfo{person}{Radu-Daniel Vatavu} {and}
  \bibinfo{person}{Jacob~O. Wobbrock}.} \bibinfo{year}{2015}\natexlab{}.
\newblock \showarticletitle{Formalizing Agreement Analysis for Elicitation
  Studies: New Measures, Significance Test, and Toolkit}. In
  \bibinfo{booktitle}{\emph{Proceedings of the 33rd Annual ACM Conference on
  Human Factors in Computing Systems}} (Seoul, Republic of Korea)
  \emph{(\bibinfo{series}{CHI ’15})}. \bibinfo{publisher}{Association for
  Computing Machinery}, \bibinfo{address}{New York, NY, USA},
  \bibinfo{pages}{1325–1334}.
\newblock
\showISBNx{9781450331456}
\urldef\tempurl%
\url{https://doi.org/10.1145/2702123.2702223}
\showDOI{\tempurl}


\bibitem[\protect\citeauthoryear{Vatavu and Wobbrock}{Vatavu and
  Wobbrock}{2016}]%
        {VAT+16}
\bibfield{author}{\bibinfo{person}{Radu-Daniel Vatavu} {and}
  \bibinfo{person}{Jacob~O. Wobbrock}.} \bibinfo{year}{2016}\natexlab{}.
\newblock \showarticletitle{Between-Subjects Elicitation Studies: Formalization
  and Tool Support}. In \bibinfo{booktitle}{\emph{Proceedings of the 2016 CHI
  Conference on Human Factors in Computing Systems}} (San Jose, California,
  USA) \emph{(\bibinfo{series}{CHI ’16})}. \bibinfo{publisher}{Association
  for Computing Machinery}, \bibinfo{address}{New York, NY, USA},
  \bibinfo{pages}{3390–3402}.
\newblock
\showISBNx{9781450333627}
\urldef\tempurl%
\url{https://doi.org/10.1145/2858036.2858228}
\showDOI{\tempurl}


\bibitem[\protect\citeauthoryear{Villarreal-Narvaez, Vanderdonckt, Vatavu, and
  Wobbrock}{Villarreal-Narvaez et~al\mbox{.}}{2020}]%
        {VIL+20}
\bibfield{author}{\bibinfo{person}{Santiago Villarreal-Narvaez},
  \bibinfo{person}{Jean Vanderdonckt}, \bibinfo{person}{Radu-Daniel Vatavu},
  {and} \bibinfo{person}{Jacob~A Wobbrock}.} \bibinfo{year}{2020}\natexlab{}.
\newblock \showarticletitle{A Systematic Review of Gesture Elicitation Studies:
  What Can We Learn from 216 Studies}. In \bibinfo{booktitle}{\emph{Proceedings
  of ACM Int. Conf. on Designing Interactive Systems (DIS'20)}}.
  \bibinfo{publisher}{ACM Press}, \bibinfo{address}{Eindhoven},
  \bibinfo{pages}{NA}.
\newblock


\bibitem[\protect\citeauthoryear{Wobbrock, Aung, Rothrock, and Myers}{Wobbrock
  et~al\mbox{.}}{2005}]%
        {WOB+05}
\bibfield{author}{\bibinfo{person}{Jacob~O Wobbrock},
  \bibinfo{person}{Htet~Htet Aung}, \bibinfo{person}{Brandon Rothrock}, {and}
  \bibinfo{person}{Brad~A Myers}.} \bibinfo{year}{2005}\natexlab{}.
\newblock \bibinfo{title}{Maximizing the guessability of symbolic input}.
\newblock
\newblock


\bibitem[\protect\citeauthoryear{Wobbrock, Morris, and Wilson}{Wobbrock
  et~al\mbox{.}}{2009}]%
        {WOB+09}
\bibfield{author}{\bibinfo{person}{Jacob~O Wobbrock},
  \bibinfo{person}{Meredith~Ringel Morris}, {and} \bibinfo{person}{Andrew~D
  Wilson}.} \bibinfo{year}{2009}\natexlab{}.
\newblock \showarticletitle{User-defined Gestures for Surface Computing}. In
  \bibinfo{booktitle}{\emph{Proceedings of the {SIGCHI} Conference on Human
  Factors in Computing Systems}} (Boston, MA, USA) \emph{(\bibinfo{series}{CHI
  '09})}. \bibinfo{publisher}{ACM}, \bibinfo{address}{New York, NY, USA},
  \bibinfo{pages}{1083--1092}.
\newblock


\bibitem[\protect\citeauthoryear{Wolf, Naumann, Rohs, and M\"{u}ller}{Wolf
  et~al\mbox{.}}{2011}]%
        {WOL+11}
\bibfield{author}{\bibinfo{person}{Katrin Wolf}, \bibinfo{person}{Anja
  Naumann}, \bibinfo{person}{Michael Rohs}, {and} \bibinfo{person}{J\"{o}rg
  M\"{u}ller}.} \bibinfo{year}{2011}\natexlab{}.
\newblock \showarticletitle{Taxonomy of Microinteractions: Defining
  Microgestures Based on Ergonomic and Scenario-dependent Requirements}. In
  \bibinfo{booktitle}{\emph{Proceedings of the 13th IFIP TC 13 International
  Conference on Human-computer Interaction - Volume Part I}} (Lisbon, Portugal)
  \emph{(\bibinfo{series}{INTERACT'11})}. \bibinfo{publisher}{Springer-Verlag},
  \bibinfo{address}{Berlin, Heidelberg}, \bibinfo{pages}{559--575}.
\newblock
\showISBNx{978-3-642-23773-7}
\urldef\tempurl%
\url{http://dl.acm.org/citation.cfm?id=2042053.2042111}
\showURL{%
\tempurl}


\bibitem[\protect\citeauthoryear{Zai{\c t}i, Pentiuc, and Vatavu}{Zai{\c t}i
  et~al\mbox{.}}{2015}]%
        {VAT+15b}
\bibfield{author}{\bibinfo{person}{Ionu{\c t}-Alexandru Zai{\c t}i},
  \bibinfo{person}{{\c S}tefan-Gheorghe Pentiuc}, {and}
  \bibinfo{person}{Radu-Daniel Vatavu}.} \bibinfo{year}{2015}\natexlab{}.
\newblock \showarticletitle{On free-hand {TV} control: experimental results on
  user-elicited gestures with Leap Motion}.
\newblock \bibinfo{journal}{\emph{Pers. Ubiquit. Comput.}}
  \bibinfo{volume}{19}, \bibinfo{number}{5} (\bibinfo{date}{Aug.}
  \bibinfo{year}{2015}), \bibinfo{pages}{821--838}.
\newblock


\end{thebibliography}

\received{July 2020}
\received[revised]{August 2020}
\received[accepted]{September 2020}

\end{document}